\documentclass[prb,aps,twocolumn,floatfix,showpacs,amsmath,amssymb]{revtex4}
\usepackage{bm}
\usepackage{graphicx}

\newcommand{\be}{\begin{equation}}
\newcommand{\ee}{\end{equation}}
\newcommand{\bea}{\begin{eqnarray}}
\newcommand{\eea}{\end{eqnarray}}
\newcommand{\barz}{\bar{z}}

\begin{document}
\def\tit#1#2#3#4#5{{#1} {\bf #2}, #3 (#4)}

\title{Spin-orbital effects in magnetized quantum wires and spin chains}
\author{Suhas Gangadharaiah, Jianmin Sun and Oleg A. Starykh}
\affiliation{Department of Physics, University of Utah, Salt Lake City, UT 84112}
\date{\today}

\begin{abstract}
We present analysis of the interacting quantum wire problem in the presence of
magnetic field and spin-orbital interaction. We show that
an interesting interplay of Zeeman and spin-orbit terms,
facilitated by the electron-electron interaction, results in the
spin-density wave (SDW) state when the magnetic field and spin-orbit
axes are {\sl orthogonal}. We show that this instability is enhanced
in a closely related problem of Heisenberg spin chain with
asymmetric uniform Dzyaloshinskii-Moriya (DM) interaction.
Magnetic field in the direction perpendicular to the DM anisotropy
axis results in staggered long-range magnetic order along the orthogonal
to the applied field direction. We  explore consequences of the uniform
DM  interaction for the electron spin resonance (ESR) measurements,
and point out that they provide way to probe right- and left-moving
excitations of the spin chain separately.
\end{abstract}
\maketitle

\section{Introduction}
Over the last several years there has been a remarkable growth
in research activity in the field of spintronics  with the
ultimate goal to  fabricate novel spin-filter devices which can
control and manipulate  the electron spins \cite{datta}. The proposals for such
a spin-filter device contains two achievable attributes, a
ballistic quantum wire and the presence  of a tunable Rashba
spin-orbit coupling responsible for controlling the electron spin.
Ballistic quantum wires are created in a 2DEG by cleaved edge over
growth, whereas the  Rashba effect arises due to  the asymmetry
associated with the confinement potential \cite{rashba}. The asymmetry and hence
the Rashba coupling strength can be  controlled  by applying the
gate voltage. Although the role of spin-orbital and  magnetic (Zeeman)
fields in the
electric and spin transport is well understood for a
non-interacting quantum wire \cite{moroz,streda,levitov,pereira,halperin},
the case of interacting electrons remains the subject of active
research \cite{kimura96,samokhin00,hausler01,iucci03,governale04,yu04,gritsev05,
hyunlee05,hikihara05}.

It should be noted that finite spin-orbit coupling is very natural, and, strictly
speaking, unavoidable, in semiconducting quantum wires due
to pronounced structural asymmetry inherent in the fabrication
process. Also, in addition to the noted asymmetry of confining potentials
(which include quantum-well potential that confines electrons to a 2D layer
as well as transverse [in-plane] potential that forms the one-dimensional
channel \cite{moroz}), spin-orbit interaction is inherent to
semiconductors of either zinc-blende or wurtzite lattice structures
lacking inversion symmetry \cite{dresselhaus}.

Another very interesting system that motivates our investigation is provided
by one-dimensional electron surface states on vicinal surface of gold \cite{vicinal-au}
as well as by electron states of self-assembled gold chains on stepped
Si$(111)$ surface of silicon \cite{himpsel06}. In both of the systems
one-dimensional ballistic channels appear due to atomic reconstruction
of surface layer of atoms, see also \cite{mugarza03,ortega05}.
The resultant surface electronic states lie within the bandgap of bulk states,
and thus, to high accuracy, are decoupled from electrons in the bulk.
Spin-orbit interaction is unexpectedly strong in these systems, with
the spin-orbit energy splitting of the order of $100$ meV.
In fact, spin-split subbands of Rashba type have been observed
in angular resolved photoemission spectroscopy (ARPES) in both
two-dimensional \cite{lashell} and one-dimensional settings \cite{vicinal-au,himpsel06}.
The very fact that the two (horizontally) spin-split parabolas are observed
in ARPES speaks for high quality and periodicity of the obtained
surface channels.

As we show below, the most interesting situation involves
electrons subjected to both spin-orbital and magnetic fields.
While it is perhaps impractical to think of ARPES measurements
in the presence of magnetic field, it is quite possible to imagine
experiments on {\sl magnetic} metal surfaces \cite{krupin05,dedkov08}.
It is then natural to investigate combined effect of non-commuting
spin-orbit and Zeeman interactions, together with electron-electron
interaction, on the one-dimensional system of electrons.

Electrons in a quantum wire (or, in a one-dimensional surface channel)
are a good  realization
for a Tomonaga-Luttinger liquid  and serves as an ideal system for
the study of the interplay of magnetic field and Rashba spin-orbit
effect on the interacting quantum wire.
The magnetic field  breaks the
time-reversal symmetry of the Hamiltonian and
splits the band of free electrons into two, corresponding to
up-spin and down-spin electrons, reducing spin-rotational symmetry of
the system  from SU(2) to U(1).
Subsequent inclusion of  the Rashba term,
 $H_R \propto \vec{\sigma}\times\vec{p}\cdot \hat{z}$,
 see Eq.\eqref{eq:HR}, breaks this U(1)
symmetry (observe that $[\sigma_z, H_R]\neq 0$).
In addition, the spin-orbit interaction (SOI) $H_R$ breaks spatial
inversion symmetry ${\cal P}: x \to -x$.
A consequence of the fully broken SU(2) symmetry is the generation
of new scattering processes which are no longer spin-conserving.
These are the Cooper scattering processes in which a pair of
electrons in the lower band scatter to the upper band and
vice versa \cite{sun07}.
 A relevant Cooper term
creates a gap in the energy spectrum leading to  a long range
spin-density wave order. We find that  the gap strength is
proportional to the backscattering ($2k_F$ component) of the
electron-electron interaction potential and the ratio of the
Rashba to the Zeeman energy. For  a large enough gap the ordering
in the spin-density wave can crucially suppress the backscattering
process of electrons from an isolated impurity.
Brief description of our main results was previously given in Ref.~\onlinecite{sun07}.

We will also analyze  an alternate  system, a Mott-Hubbard Heisenberg
spin-1/2 chain, in the presence of magnetic field and
Dzyloshinskii-Moriya (DM) interaction. A magnetic field applied at
an angle perpendicular to the DM anisotropy axis breaks the
continuous U(1) symmetry and consequently a true long range order can
develop in the spin-chain. The case of a staggered DM term has
been studied experimentally \cite{dender97} and
theoretically \cite{ao}, and has been shown to
open up a gap in the spectrum with the gap scaling as $B^{2/3}$
with the magnetic field $B$. The case of a uniform DM term
is also experimentally relevant, see for example Ref.\onlinecite{zakharov06},
but has not been discussed much theoretically.
We will show that the case of a uniform DM
term and perpendicular magnetic field can be described analogously
to the quantum wire in the presence of spin-orbit interaction and
magnetic field.

The outline for the paper is as follows: In
Sec.\ref{sec:FreeElectrons} we will review non-interacting electrons
in the presence of magnetic field and Rashba spin-orbit term. We
then consider interaction effects by using standard bosonization
approach. In Sec.\ref{sec:rg} we will perform renormalization group
analysis to determine relevant and irrelevant terms. In Sec.\ref{sec:pert}
we  use perturbative approach to generate relevant terms.  The
role of relevant terms on the transport property of electrons in
the presence of a single impurity is analyzed in section \ref{sec:imp}. In
Sec.\ref{sec:heis}, we consider a spin-$1/2$ Heisenberg antiferromagnetic
chain in the presence of magnetic field and uniform DM term  and
compare this system with the quantum wire. Comparison with
the case of a staggered
DM term is considered in Sec.\ref{sec:stagDM}.
In Section~\ref{sec:esr} we discuss the role ESR measurements
can play in unraveling the DM term in the spin chain.
Technical details of our calculations are described in Appendices.

\section{1D Electrons in the presence of magnetic field and Rashba spin orbit term}
\label{sec:FreeElectrons}

The Hamiltonian for an electron  subject to
the Rashba spin-orbit term  and in the presence of magnetic field
is given by
\begin{eqnarray}
\label{eq:H0-0}
H_0 &=& \frac{\hbar^2(p^2_x + p^2_y)}{2m}  +  V(x)
 - g\mu_B \frac{\vec{\sigma}}{2}\cdot\vec{B}  + H_R,\\
H_R & = & \frac{\alpha_R}{\hbar}(\sigma_x p_y - \sigma_y p_x)
\label{eq:HR}
\end{eqnarray}
where $\alpha_R$ is the Rashba spin-obrit coupling, $g$ is the
effective Bohr magneton, $B$ is the magnetic field, $\sigma_\mu$ ($ \mu =
x, y, z) $ are the Pauli matrices and the potential $V(x) =
m\omega^2 x^2/2$ typically confines the particle in the
$x-$direction.
When the confining potential is strong enough so that the
width  of the wire $\sqrt{\hbar/(2 m \omega)}$ is much smaller than the  electron
Fermi-wavelength, only the first sub-band  is occupied and
the Hamiltonian \eqref{eq:H0-0} acquires a one-dimensional form
\begin{eqnarray}
H_0 \approx \frac{\hbar^2 k^2}{2m}  +
\frac{\alpha_R}{\hbar}(\sigma_x k) - g\mu_B \frac{\vec{\sigma}}{2}\cdot\vec{B}  .
\label{eq:H0}
\end{eqnarray}
Here and below $k$ is electron's momentum along the axis of the wire,
which we will denote as $x$-axis in the following for notational convenience.
We also set $\hbar=1$.

It is easy to see that in the {\sl absence} of magnetic field SOI in \eqref{eq:H0} can be
easily gauged away via the {\sl spin-dependent} shift of the momentum,
$H_0 (B=0) \propto (k + m\alpha_R \sigma_x)^2$.
Corrections arising from the omitted  term
$\alpha_R \sigma_y p_x$ \eqref{eq:HR} produce small spin-dependent variations of
the velocities of right- and left-moving particles \cite{samokhin00}.
These, however, are not important for our purposes
for as long as $ \alpha_R k_F \ll E_F $,
which is the limit (along with $ \Delta_Z \ll E_F $) considered in this work. With interactions included, electrons
form Luttinger liquid with somewhat modified  critical exponents, in comparison
with the standard case of no SOI \cite{samokhin00,iucci03}.

Most interesting situation arises when both SOI and Zeeman terms are
present simultaneously and do not commute with each other as happens
when spin-orbital axis ($\sigma_x$ in \eqref{eq:H0}) is different from the
magnetic field direction. In what follows we choose  magnetic field to point
along  the $z-$direction, $\vec{B} = B \hat{z}$.
The energy eigenvalues of the Hamiltonian \eqref{eq:H0}
is found as \cite{levitov,pereira}
\begin{eqnarray}
\epsilon_{\pm} =
\frac{k^2}{2m}  \pm
\sqrt{(\alpha_R k)^2  + (\frac{\Delta_z}{2})^2   },
\label{eq:epsilon_pm}
\end{eqnarray}
and  the momentum dependent eigen-spinors are, for
$\epsilon_+(k)$:
\bea
 |\chi_{+}(k)\rangle \equiv\left[\begin{array}{cc} \sin\frac{ \gamma(k)}{2} \\
 \cos \frac{ \gamma(k)}{2}
\end{array}\right],
\label{eq:chi-plus}
\eea
\\
and for $\epsilon_-(k)$:
\bea
     |\chi_{-}(k)\rangle \equiv \left[\begin{array}{cc} \cos\frac{ \gamma(k)}{2}
\\
 -\sin \frac{ \gamma(k)}{2}
\end{array}\right],
\label{eq:chi-minus}
\eea
where $\Delta_z = g \mu_B B$ and rotation angle $\gamma(k)$ is introduced
\be
\gamma(k)= \arctan \frac{2\alpha_R k}{ \Delta_z}.
\label{eq:gamma}
\ee
Notice that particles  with momentum $\pm k$ experiences an
effective magnetic field $\vec{B}_{\rm eff} = \mp 2 \alpha_R k/ (g
\mu_B ) \hat{x} + B \hat{z} $, thus  spin directions in each band varies with the momentum.
 Going from  left to the right side of the band, the spins
"rotate" along the counter-clockwise direction. At $k=0$, the
effective magnetic field is just the applied field, thus the
separation between the two bands is minimum and the spins align
according to the applied field, i.e., along the $\mp z-$direction.
For states with the same energy
the spin states of $+$ and $-$ bands are  no longer orthogonal if
there is a finite magnetic field and Rashba spin-orbit coupling.
In particular the right and left  Fermi levels satisfy  the
following property
\be
k_-^{R/L} + k_+^{R/L} \approx \pm 2k_F
\label{eq:k_pm-sum}
\ee
and
\be \delta k_F
= |k_-^{R/L}-k_+^{R/L}| \approx \frac{m}{k_F} \sqrt{(2 k_F
\alpha_R)^2 + (\Delta_z)^2}.
\label{eq:delta-k_F}
\ee
The magnitude of
velocities at the Fermi level are
\be
|u_\pm| \approx v_F \mp  \frac{(\Delta_z)^2}
{2 k_F\sqrt{4(\alpha_R k_F)^2 +   (\Delta_z)^2}},
\label{eq:v_pm}
\ee
where $v_F = k_F/m$. The spin overlap between the upper ($+$) and lower ($-$)
band is non-zero,
\be
\langle\chi(k_+^{R/L})|\chi( k_-^{R/L})\rangle =
\sin\frac{\gamma(k_+^{R/L})-\gamma(k_-^{R/L})}{2}.
\label{eq:overlap1}
\ee
 As will be discussed
in the next section, the non-orthogonality of the spin-states
acquires important consequences when one turns on the
electron-electron interactions.

\begin{figure}[ht]
  \centering
   \includegraphics[width=2.0in]{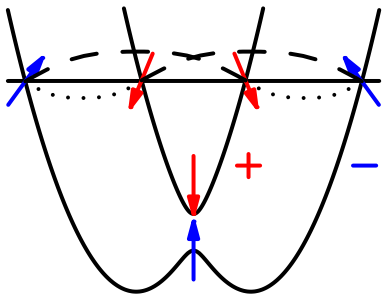}
       \caption{Occupied subbands $\epsilon_\pm$ of Eq.~\ref{eq:epsilon_pm}. Arrows illustrate spin polarization
    in different subbands. Dashed (dotted) lines indicate exchange (direct) Cooper scattering processes.}
  \label{fig:1}
\end{figure}

\section{Intersubband Interaction Effects}
\label{sec:intersubband}

The interaction part  in  terms of
the particle-field operators $\Psi_\sigma(x)$ and
$\Psi^\dagger_\sigma(x)$ ( $\sigma$ and $\sigma'$ are the spin
indices) is given by
 \be
 H_{\text{int}} =  \frac{1}{2} \int dx dx'
U(x-x') \Psi_\sigma^\dagger(x)
\Psi_{\sigma'}^\dagger(x')\Psi_{\sigma'}(x')\Psi_\sigma(x),
\label{eq:Hint}
 \ee
where the  summation  on  pairs of identical spin indices is
assumed and $U(x-x')$ is the screened (by surrounding gates) interaction between the
electrons. The field $\Psi_\sigma(x)$ is conventionally defined in
terms of the annihilation operator $a_{\sigma}(k)$  of a free
particle in the state $e^{i k x}$ and with spin $\sigma=\uparrow,\downarrow$: $
\Psi_\sigma(x) = \int \frac{dk}{2\pi} e^{i k x} a_\sigma(k)$.
Alternatively,
annihilation operators $a_{\pm}(k)$ of particles which are the eigen-states  of
the Hamiltonian (\ref{eq:H0-0}) with  eigen-energies
$\epsilon_{\pm}(k)$,  can be used to represent the
field operator as follows:
\be \Psi_\sigma(x) =
\sum_{\nu=\pm}\int\frac{dk}{2\pi} e^{i k x}
\langle\chi_{\nu}(k)|\sigma\rangle a_{\nu}(k).
\label{eq:psi-1}
\ee
The low energy
physics of the interacting wire is described by linearizing the
spectrum near the Fermi-points, $\pm k_\pm$.
The $\sigma=\uparrow,\downarrow$ field operators
are now  described, in coordinate space,
in terms of the chiral right ($R_\nu$) and left ($L_\nu$) movers of $\nu=\mp$ subbands
as follows
 \be \Psi_\sigma(x) =
\sum_{\nu=\mp} \langle \chi_\nu(k_\nu)| \sigma\rangle
 e^{ik_{\nu}x} R_{\nu} + \langle \chi_\nu(-k_\nu)| \sigma\rangle e^{-ik_{\nu}x} L_{\nu}.
 \label{eq:3}
\ee Following \cite{review}, we decompose the interaction part of the Hamiltonian
(\ref{eq:Hint}) into  intra-subband $H_{\text{intra}}$ and inter-subband
$H_{\text{inter}}$ scattering processes. The intra-subband process,
$H_{\text{intra}}$,  describes the interaction  between electrons in the
same subband and involves the standard forward and backscattering
processes, see Appendix~\ref{sec:ap-intra}.  The second scattering mechanism, $H_{\text{inter}}$, involves
scattering between electrons in  different subbands and  can be
conveniently divided into {\sl forward}, {\sl backward} and {\sl Cooper} scattering processes.
Below we will discuss  the inter-subband scattering processes in more detail.
The forward scattering
process involves interaction between $q\approx 0$ components of the densities
in the two subbands
\begin{eqnarray}
&& H_{\text{inter}}^{\text{F}} = \frac{1}{2} U(0)\int dx
\sum_{\nu=\pm}(R^\dagger_\nu R_\nu + L^\dagger_\nu L_\nu)\nonumber\\ &&
\times (R^\dagger_{-\nu} R_{-\nu} + L^\dagger_{-\nu} L_{-\nu}).
\label{eq:inter-FS}
\end{eqnarray}
The inter-subband backscattering process is classified into {\sl direct} and
{\sl exchange} scattering. Direct backscattering process
involves $q \approx 2 k_\pm$ components of the densities in
the two subbands:
a left (right) moving electron in the subband $\nu$ ($-\nu$)
changes its direction to become a right (left)
moving one while remaining in the same band,
\begin{eqnarray}
&&H_{\text{inter}}^{\text{d-B}} =
\cos[\gamma(k_+)] \cos[\gamma(k_-)] U(k_+ + k_{-} )
{} \nonumber\\
 && \times \sum_{\nu=\pm}\int dx ~e^{i2(k_\nu - k_{-\nu})x}
 (R^\dagger_\nu L_\nu)(L^\dagger_{-\nu} R_{-\nu}) .
 \label{eq:inter-dB}
\end{eqnarray}
This contribution
involves an oscillatory factor $\exp[i 2 (k_\nu -k_{-\nu})x]$ in the
integral due to the non-conservation of momentum during the scattering.

The other backscattering process is via a momentum conserving {\sl exchange}
mechanism, where electrons again scatter by large momentum transfer and in
the process {\sl exchange their bands}. The scattering channel  conserves momentum
and reads
\begin{eqnarray}
&&H_{\text{inter}}^{\text{ex-B}} = \frac{1}{2}\sum_{\nu=\pm}\Big\{U(k_\nu -
k_{-\nu})\sin^2[\frac{\gamma(k_\nu)- \gamma(k_{-\nu}) }{2}]
\nonumber\\
&& \times \int dx \Big((R^\dagger_\nu
R_{-\nu})(R^\dagger_{-\nu} R_{\nu}) + (L^\dagger_\nu
L_{-\nu})(L^\dagger_{-\nu} L_{\nu})\Big){}
\nonumber\\
 && +  2 U(k_\nu +
k_{-\nu})\sin^2[\frac{\gamma(k_\nu) + \gamma(k_{-\nu}) }{2}]
\nonumber\\
&& \times \int dx (R^\dagger_\nu L_{-\nu})(L^\dagger_{-\nu} R_{\nu}) \Big\}.
\label{eq:inter-exB}
\end{eqnarray}
Note the appearance of (squared) wave function overlap factors,
$\propto \langle \chi(k_\nu) | \chi(\pm k_{-\nu})\rangle $,
which signify the exchange nature of the scattering.
Note also that these factors are non-zero
due to a finite Rasbha coupling $\alpha_R$, which allows  electrons to
scatter without conserving their spins.

The  Cooper scattering process, which is
central to our story, involves
scattering of a pair of opposite movers (right and left) in the
subband $\nu$ into a similar pair in the other, $-\nu$, subband.
Each pair has zero total momentum which remains conserved in this scattering.
Being  {\sl pair-tunneling} like, Cooper scattering requires
{\sl non-conservation} of spin. It represents, for example,
a scattering of a pair of two (almost) down-spin electrons into a pair
of two (almost) up-spin ones.
The Cooper scattering reads
\begin{eqnarray}
&&H_{\text{inter}}^{\text{C}} =
\int dx \Big\{U(k_{-} - k_{+})\sin^2[\frac{\gamma(k_{-})- \gamma(k_{+}) }{2}]
\nonumber\\
&& - U(k_{-} + k_{+}) \sin^2[\frac{\gamma(k_{-}) + \gamma(k_{+})}{2}] \Big\}
\nonumber\\
&& \times   (R^\dagger_{-} L^\dagger_{-} R_{+} L_{+} + \text{h.c.}).
\label{eq:inter-Cooper}
\end{eqnarray}
The first term ({\sl direct} Cooper scattering) is due to electrons in the band $R_\nu$
 and $L_\nu$ jumping into $R_{-\nu}$ and $L_{-\nu}$, respectively.
 The coefficient for this term is $\propto U(\delta k_F)
 \sin^2[(\gamma_--\gamma_+)/2]$, where $\delta k_F = |k_\nu - k_{-\nu}|$ is
 the momentum transfer for an electron and
 $\sin^2[(\gamma_{-} - \gamma_+)/2]$ is the squared  overlap integral.
 (For brevity, we denote $\gamma(k_\pm) = \gamma_\pm$ here and in the following.)
 The second process ({\sl exchange}) with electron scattering from  $R_\nu$
 and $L_\nu$ to $L_{-\nu}$ and $R_{-\nu}$, respectively, involves
 a coefficient, $\propto U(k_\nu + k_{-\nu})
 \sin^2[(\gamma_-+\gamma_+)/2]$, with a larger, $\sin^2[(\gamma_-+\gamma_+)/2]$,
  overlap integral. The bigger overlap for this second ($R_\nu \leftrightarrow L_{-\nu}$)
  process is also rather clear from pictorial representation of spin orientation
  in different subbands, as shown in Fig.~\ref{fig:1}.
  For the case of short-ranged (screened) interaction potential a simple
  estimate, using (\ref{eq:k_pm-sum}) and (\ref{eq:delta-k_F}),
  \be
  \frac{U(k_{-} - k_{+})\sin^2[\frac{\gamma_{-}- \gamma_{+} }{2}]}
  {U(k_{-} + k_{+}) \sin^2[\frac{\gamma_{-} + \gamma_{+}}{2}]} \approx \frac{U(\delta  k_F)}{U(2 k_F)} \Big(\frac{\Delta_z}{2 E_F}\Big)^2 \ll 1
  \ee
  shows that the second, {\sl exchange} Cooper process, dominates.
This defines the regime to be considered in this work.

\begin{figure}[ht]
  \centering
   \includegraphics[width=2.5in]{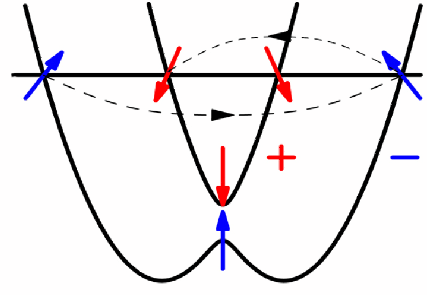}
       \caption{Asymmetric back-scattering processes.}
  \label{fig:asym}
\end{figure}
Finally we take into account two classes of momentum non-conserving scattering processes
where one of them exhibits  mixed features of Cooper and back-scattering
(called as asymmetric back-scattering process) while the other one has  features reminiscent of Cooper
and forward scattering processes ( asymmetric forward-scattering
process). A typical asymmetric back (forward) scattering event
involves total momentum change $\pm \delta k_F$, see Fig.\ref{fig:asym}.
For example, right and left moving fermions in the
same sub-band  scatter into left (right) and right (left)
moving  fermions respectively, with one of the fermions now
in a different sub-band. Alternatively, the oppositely moving
fermions may be in different sub-bands to begin with but end up
in the same sub-band with opposite (same) momentums. The scattering process
acquires a slowly (compared to direct back scattering (\ref{eq:inter-dB})) oscillating
factor $\exp[i(k_\nu-k_{-\nu})x]$ owing to the non-conservation of momentum and is
given by
\begin{eqnarray}
&&~~~~~~H_{\text{asymm}} = \sin(\frac{\gamma_{+}+\gamma_{-}}{2})
\sum_{\nu=\pm}\text{sgn}(\nu)
U(2k_F)\cos(\gamma_{\nu})\times \nonumber\\
&&
 \int dx [e^{i(k_\nu-k_{-\nu})x}  (R^{\dagger}_{\nu}L^{\dagger}_{\nu}R_{\nu}L_{-\nu}
 -   R^{\dagger}_{-\nu}L^{\dagger}_{\nu}R_{\nu}L_{\nu}       )
 + \text{h.c.}].\label{eq:Hasymm}
\end{eqnarray}
The above expression  reflects contributions  from only the asymmetric back-scattering processes.
The asymmetric forward-scattering processes involve identical fermion operators as in
Eq.(\ref{eq:Hasymm}). However the ratio of amplitudes for asymmetric forward to asymmetric
back-scattering process is small,
\begin{eqnarray}
\frac{U(\delta k_F) \sin[\frac{\gamma_{\nu}-\gamma_{-\nu}}{2}]}{U(2k_F)
\sin{[\frac{\gamma_\nu+\gamma_{-\nu}}{2}]}\cos{\gamma_\nu}   }\approx
\frac{U(\delta k_F)}{U(2k_F)}\frac{\sqrt{\Delta_z^2 + 4 (\alpha_R k_F)^2}}{2 E_F}\ll 1, \nonumber
\end{eqnarray}
which allows us to neglect contributions from asymmetric forward-scattering processes altogether.

The electron density in
the quantum wire is assumed to be incommensurate with the lattice
spacing, hence  the Umklapp scattering process is not considered.
To summarize,  the interaction part of the Hamiltonian has been decomposed in terms
of three broadly defined scattering processes,  intra subband,  inter subband and
asymmetric scattering process,
\begin{eqnarray}
&& H = H_{\text{intra}} + H_{\text{inter}} + H_{\text{Asymm}}.\label{eq:split_hamil}
\end{eqnarray}

\subsection{Bosonization}
\label{sec:bosonization}

Bosonization is performed by expressing the  fermionic operators
in the Hamiltonian via  the chiral bosonic $\phi_{R/L_\nu}$
fields \cite{gnt-book,giamarchi-book,hubbard-ladder}.
The fermionic fields in terms of the chiral bosonic field are as follows
\begin{eqnarray}
R_{\pm} = \frac{\eta_{\pm}}{\sqrt{2\pi a_0}}  e^{i\sqrt{4\pi}
\phi_{R\pm} },\ \
 L_{\pm} = \frac{\eta_{\pm}}{\sqrt{2\pi a_0}}
e^{-i\sqrt{4\pi} \phi_{L\pm} },
\label{eq:bos1}
\end{eqnarray}
where $a_0\sim k_F^{-1}$  is the short distance cutoff and
$\eta_{\pm}$ are the Klein factors which are introduced to ensure
the correct anti-commutation relations for the fermionic
operators from different ($\pm$) subbands.
The bosonic operators obey the following commutation
relations:
\begin{eqnarray}
[\phi_{R\nu},\phi_{L\nu'}] = \frac{i}{4} \delta_{\nu \nu'}; \
\ \ \text{where} \ \ \ \nu,\nu' = \pm
\label{eq:bos2}
\end{eqnarray}
\be
 [\phi_{R/L\nu}(x),\phi_{R/L\nu'}(y)] = \pm \frac{i }{4}\delta_{\nu\nu'}\text{sign} (x-y),
 \label{eq:bos3}
 \ee
 the first of which, (\ref{eq:bos2}), ensures anticommutation between right and left
 movers from the {\sl same} subband, while the second is needed for the
 anticommutation between like species (i.e. right with right, left with left).
 Klein factors anticommute
\begin{eqnarray}
\{\eta_{\nu},\eta_{\nu'}\} = 2 \delta_{\nu \nu'}~,~
 \eta_{\nu}^\dagger = \eta_{\nu}.
 \label{eq:klein}
\end{eqnarray}
In the following we choose the gauge where $\eta_{+} \eta_{-} = i$.
The chiral $\phi_{R/L\nu}$ are expressed in terms of $\phi_\nu$
and its dual $\theta_\nu$ as follows
\begin{eqnarray}
 \phi_{R\nu}=\frac{\phi_\nu -\theta_\nu}{2}; \ \
\phi_{L\nu}=\frac{\phi_\nu + \theta_\nu}{2},
\label{eq:bos4}
\end{eqnarray}
The bosonized form  of the   Hamiltonian  is
obtained by making   use of
equations (\ref{eq:bos1}) through (\ref{eq:bos4}),
as well as the following results for (chiral) densities
\begin{eqnarray}
&& R^\dagger_\nu  R_\nu = \frac{\partial_x
\phi_{R\nu}}{\sqrt{\pi}} = \frac{\partial_x (\phi_\nu
-\theta_\nu)}{\sqrt{4\pi}}, {}\nonumber\\ && L^\dagger_\nu L_\nu =
\frac{\partial_x \phi_{L\nu}}{\sqrt{\pi}}= \frac{\partial_x
(\phi_\nu + \theta_\nu)}{\sqrt{4\pi}}
\label{eq:bos-density}
\end{eqnarray}
The (bosonized) Hamiltonian in terms of $\phi_\nu$ and $\theta_\nu$
is the sum of intra-subband, inter-subband scattering
and asymmetric scattering processes,
\begin{eqnarray}
&& H = H_{\text{intra}} + H_{\text{inter}} + H_{\text{asymm}},
\end{eqnarray}
where the intra-subband part has the usual form (see Appendix~\ref{sec:ap-intra})
\begin{eqnarray}
&&  H_{\text{intra}} = \frac{1}{2}\sum_{\nu=\pm}\int dx \Big\{v_F
(\partial_x  \theta_\nu)^2 {}\nonumber\\ && + (v_F +
\frac{U(0)-\cos^2[\gamma(k_\nu)] U(2k_\nu)}{\pi})(\partial_x
\phi_\nu)^2  \Big\}.
\label{eq:Hintra}
\end{eqnarray}
The inter-subband part of the Hamiltonian, $H_{inter}$, is given by
\be
H_{\text{inter}} = H_{\text{inter}}^{\text{F}} + H_{\text{inter}}^{\text{d-B}} + H_{\text{inter}}^{\text{ex-B}} +
H_{\text{inter}}^{\text{C}},
\ee
where
\begin{eqnarray}
&& H_{\text{inter}}^{\text{F}}  =  \frac{U(0)}{\pi} \int dx  ~\partial_x
\phi_+ \partial_x\phi_-, {} \nonumber\\
&& H_{\text{inter}}^{\text{d-B}} =  \frac{U(2k_F) \cos^2 [\gamma_F]}{2(\pi
 a_0)^2} {} \nonumber\\
&&\times\int dx  \cos [
2\sqrt{\pi}(\phi_--\phi_+) + 2 (k_+-k_-)x],{} \nonumber\\
&&
H_{\text{inter}}^{\text{ex-B}}   = - \frac{U(2k_F)}{2\pi} \sin^2[\gamma_F] {}\nonumber\\
&& \times\int dx~(\partial_x \phi_+
\partial_x \phi_- - \partial_x \theta_+ \partial_x \theta_- ), {}
\nonumber\\
  && H_{\text{inter}}^{\text{C}} =   \frac{U(2k_F)
  \sin^2[\gamma_F]}{2(\pi a_0)^2} {}\nonumber\\ &&
\times\int dx \cos [2\sqrt{\pi}(\theta_--\theta_+)].
\label{eq:Hinterband}
\end{eqnarray}
The asymmetric part has the following bosonized form,
\begin{eqnarray}
&& H_{\text{asymm}} = -\frac{\sqrt{2}U(2k_F)\sin(2\gamma_F)}{(2\pi)^{3/2}a_0}
\int dx \{\partial_x (\phi_{R-} - \phi_{R+}) {}\nonumber\\ && \times
\sin [\sqrt{4\pi}(\phi_{L-} - \phi_{L+})-\delta k_F x]
 + \partial_x (\phi_{L-} - \phi_{L+})
 {}\nonumber\\ && \times
 \sin [\sqrt{4\pi}(\phi_{R-} - \phi_{R+})-\delta k_F x]\}.
\label{eq:Hbos_asymm}
\end{eqnarray}
In deriving Eqs.~(\ref{eq:Hinterband}) and  (\ref{eq:Hbos_asymm}) we took limits $\Delta_z \ll E_F = v_F k_F$
and $\alpha_R k_F \ll E_F$ which allowed us to neglect velocity differences
(\ref{eq:v_pm}) in the two subbands and approximate $U(2k_\pm),U(\frac{3k_\nu-k_{-\nu}}{2}) \approx U(2k_F)$
and $\gamma(k_\pm) \approx \gamma(k_F) \equiv \gamma_F$.
An important  exception to this replacement is provided by $H_{\text{inter}}^{\text{d-B}}$ and $H_{\text{asymm}}$
in (\ref{eq:Hinterband}) and (\ref{eq:Hbos_asymm}), respectively, where momentum {\sl mismatch}
factors $2 (k_{-} - k_{+})x = 2\delta k_F x$ and $ (k_{-} - k_{+})x = \delta k_F x$
must be preserved. It is worth noting here that the approximations assumed
do not restrict the ratio $ 2\alpha_R k_F/(g\mu_B B) = E_{\text{s-o}}/\Delta_z$, which
can still take on any value.

A more standard representation of the Hamiltonian is in terms of
the symmetric $\phi_\rho$, $\theta_\rho$ ({\sl charge}) and
anti-symmetric $\phi_\sigma$, $\theta_\sigma$ ({\sl spin}) modes. These
combinations are defined as follows
\begin{eqnarray}
&&  \varphi_\rho = \frac{\phi_- + \phi_+}{\sqrt{2}},
\ \ \varphi_\sigma = \frac{\phi_- - \phi_+}{\sqrt{2}}, {} \nonumber\\
&&\theta_\rho = \frac{\theta_- + \theta_+}{\sqrt{2}}, \ \ \
\theta_\sigma = \frac{\theta_- - \theta_+}{\sqrt{2}}.
\label{eq:bos5}
\end{eqnarray}
The Hamiltonian  now reads
\begin{eqnarray}
&&H = H_\rho  + H_\sigma  .
\end{eqnarray}
The charge part of the Hamiltonian is harmonic
\begin{eqnarray}
 && H_\rho = \frac{1}{2}\int
dx \Big[ u_\rho K_\rho ( \partial_x \theta_\rho)^2 +
\frac{u_\rho}{K_\rho} ( \partial_x \varphi_\rho)^2 \Big].
\label{eq:H-rho}
\end{eqnarray}
The spin part is the sum of quadratic and non-linear terms,
$H_\sigma = H_\sigma^0 + H_\sigma^{\text{C}} + H_\sigma^{\text{B}}+ H_\sigma^{\text{A}}$, where
\bea
H_\sigma^0 &=& \frac{1}{2}\int dx
   \Big[  u_\sigma K_\sigma ( \partial_x \theta_\sigma)^2 +
\frac{u_\sigma}{K_\sigma} ( \partial_x \varphi_\sigma)^2 \Big] \label{eq:H-sigma-0}\\
H_\sigma^{\text{C}} &=&  \frac{U(2k_F) \sin^2 [\gamma_F]}{2(\pi a_0)^2} \int dx~ \cos [\sqrt{8\pi}\theta_\sigma] \label{eq:H-sigma-C}\\
H_\sigma^{\text{B}} &=&  \frac{U(2k_F) \cos^2 [\gamma_F]}{2(\pi a_0)^2} \int dx
\cos [\sqrt{8\pi}\varphi_\sigma  +   2 \delta k_F x]~~~~
\label{eq:H-sigma-B}\\
H_\sigma^{\text{A}} &=& H_{\text{asymm}}=i\eta_{+}\eta_{-}\frac{U(2k_F)\sin(2\gamma_F)}{(2\pi)^{3/2}a_0}{}\nonumber\\ &&
\int dx \{\partial_x (\phi_\sigma-\theta_\sigma)  \sin [\sqrt{2\pi}(\phi_\sigma +\theta_\sigma)-\delta k_F x]{}\nonumber\\&&
 + \partial_x (\phi_\sigma+\theta_\sigma) \sin [\sqrt{2\pi}(\phi_\sigma-\theta_\sigma)-\delta k_F x]\}.~~~~\label{eq:H-sigma-A}
\eea
For completeness, it is worth noting that the leading correction to these equations
is represented by the {\sl inter-mode} term
\be
H'_{\rho-\sigma} = \frac{(u_{+} - u_{-})}{2}\int dx ~( \partial_x \varphi_\rho
\partial_x  \varphi_\sigma + \partial_x \theta_\rho \partial_x \theta_\sigma) ,
\ee
which couples spin and charge sectors. Its small amplitude $(u_{+} - u_{-}) \propto \Delta_z/E_F$,
see (\ref{eq:v_pm}), justifies its neglect in the following.

Competing nature of interacting problem is clear from the presence
of two non-linear terms, \eqref{eq:H-sigma-C} and \eqref{eq:H-sigma-B},
involving non-commuting (dual) boson fields $\theta_\sigma$ and $\varphi_\sigma$,
in the Hamiltonian. Similar situation happens in models of organic conductors,
where spin-nonconserving  spin-orbit and
dipole-dipole interactions play an important role \cite{giamarchi88}.

The Luttinger liquid parameters, $K_{\rho/\sigma}$,  and charge/spin
velocities $u_{\rho/\sigma}$ are found by adding contributions from harmonic
Hamiltonians (\ref{eq:Hintra}) and $H_{\text{inter}}^{\text{F}}$ and $H_{\text{inter}}^{\text{ex-B}}$
from (\ref{eq:Hinterband}), with the result
\begin{eqnarray}
&&u_\rho \approx   v_F(1 + \frac{2U(0)-U(2k_F)\cos^2[\gamma_F]}{2\pi v_F }) ,
\nonumber\\
&& u_\sigma \approx   v_F (1 - \frac{U(2k_F)\cos^2[\gamma_F]}{2\pi v_F }) , \nonumber\\
&&K_\rho \approx  1-  \frac{2U(0) - U(2k_F)}{2\pi v_F} \leq 1 ,
\nonumber\\
&&K_\sigma \approx 1+ \frac{ U(2k_F) \cos [2 \gamma_F]}{2\pi v_F} .
\label{eq:LLparameters}
\end{eqnarray}
These expressions are perturbative in small parameters $U(0)/v_F$ and $U(2k_F)/v_F$.
Note also that physically reasonable interactions are characterized by
$U(2k_F) \leq U(0)$, where the equality sign is obtained in the limit of
fully screened, delta-function
like contact interaction between electrons.

Noting that $\gamma_F$ varies from $0$ to $\pi/2$ as the ratio $E_{\text{s-o}}/\Delta_z$
varies from $0$ to $\infty$,
\be
\gamma_F= \arctan \frac{2\alpha_R k_F}{ \Delta_z} \to \left\{
\begin{array}{cc}
0 & {\text{for}}~ E_{\text{s-o}} \ll \Delta_z \\
\pi/2 & {\text{for}}~ E_{\text{s-o}} \gg \Delta_z
\end{array}\right.,
\label{eq:gamma_F}
\ee
we observe that spin stiffness $K_\sigma$ in (\ref{eq:LLparameters})
varies from its standard value slightly {\sl above} $1$, $K_\sigma(\gamma_F \to 0) = 1 + U(2k_F)/(2\pi v_F)$,
to the value {\sl below} $1$, $K_\sigma(\gamma_F \to \pi/2) = 1 - U(2k_F)/(2\pi v_F)$.
This unusual behavior, consequences
of which are  discussed below, is rooted in the spin-orbit-broken
spin-rotational invariance of the problem, as discussed in the Introduction.

\subsection{Renormalization Group Analysis}
\label{sec:rg}
The fate of the  three non-linear terms, Cooper (\ref{eq:H-sigma-C}), backscattering (\ref{eq:H-sigma-B})  and asymmetric  (\ref{eq:H-sigma-A}),
are determined by  renormalization group (RG) analysis. The analysis is significantly
simplified by expressing the Hamiltonian in terms of current operators.
To this end, we write the Hamiltonian
in terms of  the  right and left
spin ($\vec{J}_R,\vec{J}_L$) and charge ($J_R^c, J_L^c$) currents, which
obey Kac-Moody  algebra \cite{gnt-book}. The uniform part of the
currents are expressed in terms of the chiral  right and left
moving fermions: the charge currents are
\begin{eqnarray}
&& J_{R}^c = \sum_{\nu=\mp}R^\dagger_{\nu}R_{\nu},
~~~  J_{L}^c = \sum_{\nu=\mp}L^\dagger_{\nu}L_{\nu},
\end{eqnarray}
and the spin-currents are
\begin{eqnarray}
&& \vec{J}_{R} = \sum_{\nu\nu'=\mp} R^\dagger_{\nu}\frac{\vec{\sigma}_{\nu\nu'}}{2}R_{\nu'},
\vec{J}_{L} = \sum_{\nu\nu'=\mp} L^\dagger_{\nu}\frac{\vec{\sigma}_{\nu\nu'}}{2}L_{\nu'}.
\label{eq:spin-currents}
\end{eqnarray}
As an example, the $z-$component of the right moving spin current is defined as
$J^{z}_R = (R^\dagger_{-}R_{-}-R^\dagger_{+}R_{+})/2 $. Note that in the asymptotic
limit of $\alpha_R k_F/ \Delta_Z \rightarrow 0$ the $(-,+$)
bands correspond to $(\uparrow, \downarrow$) spin bands and we recover the canonical
definition  for the spin current.

Since the charge part of the Hamiltonian is quadratic, (\ref{eq:H-rho}), and is decoupled from the spin part, it suffices to consider the RG flow of the spin part only,
$H_\sigma = H_\sigma^0 + H_\sigma^{\text{C}} + H_\sigma^{\text{B}}+ H_\sigma^{\text{A}}$. In terms of current operators it reads:
\begin{eqnarray}
 H_\sigma^0 &=& 2\pi u_\sigma \int dx [(J^z_R J^z_R + J^z_L J^z_L) - y_\sigma J^z_RJ^z_L] ,
 \nonumber\\
 H_\sigma^A &=& \pi u_\sigma y_\sigma^A
 \int dx [e^{-i\delta k_F x} (J_R^zJ_L^{-}- J_R^{+}J_L^{z}) + \text{h.c.}],
\nonumber\\
H_\sigma^B &=& \pi u_\sigma y_\sigma^B\int dx [e^{-i2\delta k_F x}J_R^{-}J_L^{+}+ \text{h.c.}],
\nonumber\\
H_\sigma^C &=&  \pi u_\sigma y_\sigma^C\int dx (J_R^{-} J_L^{-}+\text{h.c.}).
\label{eq:H-sigma-current1}
\end{eqnarray}
The initial values $(\ell=0)$ of the interaction parameters are
\begin{eqnarray}
&&y_\sigma(0)=2(K_\sigma-1) = U(2k_F) \cos[2\gamma_F]/\pi v_F,{}\nonumber\\
&&y_\sigma^C(0) =U(2k_F) \sin^2(\gamma_{F} )/\pi u_\sigma, {}\nonumber\\
&&y_\sigma^B(0)= - U(2k_F)\cos^2(\gamma_F) / \pi u_\sigma, {}\nonumber\\
&& y_\sigma^A(0) =  U(2k_F)\sin(2\gamma_F)/\pi u_\sigma.
\label{eq:y-initial}
\end{eqnarray}
Note that to first order in $U(2k_F)$ there is no difference between
$v_F$ and $u_\sigma$ in denominators of the above expressions.

It is convenient to start with formal but useful limit of $\delta k_F=0$,
where $H_\sigma$ can be compactly written in terms of spin currents
\begin{eqnarray}
&&H_\sigma(\delta k_F =0) = 2\pi u_\sigma \int dx [(J^z_R J^z_R + J^z_L J^z_L)+ {}\nonumber\\
&&+\sum_{a=x,y,z} y_a J_R^aJ_R^a
 + y_A (J_R^zJ_L^x - J_R^xJ_L^z)].
\label{eq:H-sigma-current2}
 \end{eqnarray}
Here  $y_x =y_\sigma^B+y_\sigma^C $, $y_y= y_\sigma^B-y_\sigma^C $, $y_z =-y_\sigma $ and $y_A = y_\sigma^A$.  The RG equations for the dimensionless couplings $y_{a=x,y,z,A}$ are
 easy to derive with the help of OPE (operator product expansion) technique,
\begin{eqnarray}
\frac{dy_x}{d\ell} &=& y_y y_z,\nonumber\\
\frac{dy_y}{d\ell} &=& y_z y_x + y_{A}^2 ,\nonumber\\
\frac{dy_z}{d\ell} &=& y_y y_x,\nonumber\\
\frac{dy_A}{d\ell} &=& y_y y_{A}.
\label{eq:rg-y}
\end{eqnarray}
Despite complicated appearance, the solution of this system of equations is easy.
One finds that $y_x(\ell) = y_z(\ell)$, $y_A(\ell) = - \sin[2\gamma_F]y_y(\ell)$, and
$y_x(\ell) = \cos[2\gamma_F] y_y(\ell)$. As a result, the system is reduced to a single
equation $dy_y/d\ell=y_y^2$, solution of which is standard:
$y_y(\ell) = y_y(0)/(1-y_y(0)\ell) \to -1/\ell$ for $\ell\to\infty$.

Thus, in the absence of momentum mismatch $\delta k_F$ between the two subbands,
{\em all} perturbations in \eqref{eq:H-sigma-current2} are marginally irrelevant and
logarithmically decay to zero. This simply reflects rotational $SU(2)$ symmetry of the
problem in the absence of spin-orbit and Zeeman fields.

\begin{figure}[ht]
  \centering
   \includegraphics[width=2.0in]{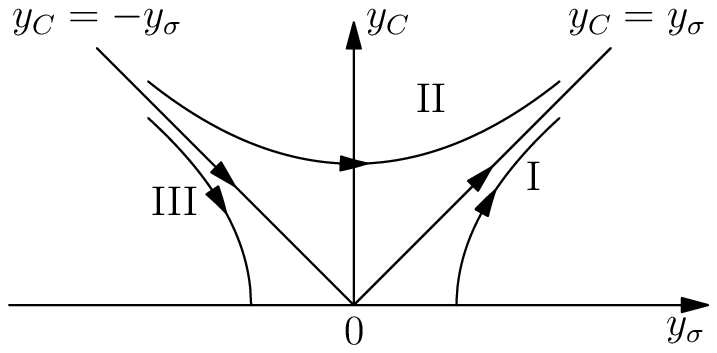}
       \caption{RG flow of \eqref{eq:rg1}.}
  \label{fig:rgflow}
\end{figure}

The full problem, with $\delta k_F\neq 0$, is solved by neglecting all momentum
non-conserving terms in $H_\sigma$: being marginal in the $\delta k_F=0$ limit,
such terms become infinitely irrelevant for  $\delta k_F\neq 0$.
More carefully, we can follow full RG \eqref{eq:rg-y} until
$\ell_z = \ln[1/(a_0 \delta k_F)] = \ln[k_F/\delta k_F]$ is reached --
beyond this scale oscillating terms average to zero. The end result is that we
are allowed to disregard $H_\sigma^{A,B}$ terms in \eqref{eq:H-sigma-current1}.
RG equations for remaining couplings can be obtained from \eqref{eq:rg-y} by nullifying all
but two, $y_\sigma$ and $y_\sigma^C \equiv y_C$, couplings.
This leads to the standard
system of two KT equations (see, for example, Ref.\onlinecite{gnt-book, giamarchi-book}).
\be
\frac{dy_\sigma}{d\ell} = y_C^2, ~~~\frac{dy_C }{d\ell} = y_{\sigma} y_C.
\label{eq:rg1}
\ee
Note that initial values of these couplings are given by corresponding solutions
of \eqref{eq:rg-y} evaluated at  $\ell = \ell_z$.

Solution of this system is determined by the integral of motion
$\mu^2 = y_C^2 - y_\sigma^2$ and
the ratio of the initial couplings $y_\sigma(0)/y_C(0) = -\cos[\alpha]$.
In terms of these parameters it reads \cite{gnt-book}
\be
y_\sigma(\ell) = - \mu \cot[\mu \ell + \alpha],
~~~y_C(\ell) = \mu/\sin[\mu \ell + \alpha].
\label{eq:rg1-solution}
\ee
There are three different regimes, illustrated in Figure~\ref{fig:rgflow}.

{\bf I}, strong coupling: $1/\sqrt{3} \geq \sin[\gamma_F] \geq 0$.
Here $\mu = -i m$, with $m>0$, and $\alpha = \pi + i \beta$, with
$\beta > 0$. In this regime both coupling flow to strong coupling,
reaching pole singularity at $\ell_0 = \beta/m$.

{\bf II}, cross-over regime: $1 \geq \sin[\gamma_F] \geq 1/\sqrt{3}$.
Here $\mu > 0$ and $0 \leq \alpha \leq \pi$. The flow is still to
strong coupling, but via an intermediate (cross-over) region
(for $0 \leq \alpha \leq \pi/2$) where $y_C(\ell)$ initially decreases.
Eventually both $y_{\sigma,C}$ reach strong coupling, at
$\ell_0 = (\pi - \alpha)/\mu$.

{\bf III}, weak coupling: This obtains when $\mu = i m$, $m > 0$, and
$\alpha = i \beta$, with $\beta > 0$. In this situation $y_\sigma \to - m$
as $\ell \to \infty$, while $y_C \to 0$. This is critical (Luttinger liquid)
phase of the spin sector. It is, however, \emph{not} realized in our problem
as the requirement $ -y_\sigma(0) > y_C(0) > 0$ is
equivalent to $\sin[\gamma_F] > 1$ which is clearly not possible.

The conclusion is then that Cooper phase is realized for arbitrary
value of $\pi/2 > \gamma_F \geq 0$, i.e. for arbitrary ratio
of SO to Zeeman energies, $\tan[\gamma_F] = 2\alpha_R k_F/\Delta_z$.
This finding of the Cooper phase, which has the meaning of the spin-orbit stabilized
{\sl spin-density-wave} (SDW$_x$) phase (see next Section),
constitutes the main result of our work.
We have previously discussed the limit
of small $\gamma_F$, which is physically most
transparent, in Ref.~\onlinecite{sun07}.

\subsection{The nature of the Cooper ordering}
\label{sec:nature-C}

We now consider the physical meaning of the Cooper instability.
For simplicity we focus on the regime {\bf I} of the previous Section.
Being relevant, the Cooper term (\ref{eq:H-sigma-C}) grows
in magnitude and reaches strong coupling limit when $y_C(\ell_c) \sim 1$ while
$K_\sigma \to 2$ \cite{gnt-book}.
A positive value of $g_C$ results in
$\theta_\sigma$  field being pinned to one of
the semi-classical minima $\theta_\sigma^{\text{cl}}= (m+\frac{1}{2}) \sqrt{\pi/2}$ ($m\in Z$).
The energy cost of (massive) fluctuations $\delta \theta_\sigma$ near these minima
represents  spin gap which can be estimated as
\be
\Delta_c \approx  \frac{v_F}{\xi} = v_F \Big( \frac{\gamma_F^2 U(2k_F)}
{\pi v_F}\Big)^{K_\sigma/2(K_\sigma-1))}.
\label{eq:Delta-c}
\ee
Here $\xi = a_0 e^{\ell_c}$ is the correlation length,
$\xi \sim [\pi v_F/(U(2k_F) \gamma_F^2)]^{K_\sigma/(2(K_\sigma-1))}$.
Physical meaning of these minima follows from the analysis of spin correlations.

We start with spin density $S^a = \Psi^\dagger_s \sigma^a_{s,s'} \Psi_{s'} /2$, which
is defined with respect to the standard spin basis, $s=\uparrow,\downarrow$.
We focus on ``$2k_F$"-components of spin, where quotation marks are used to
remind that large-momentum components of spin density include contributions from both
$k_{+} + k_{-} = 2k_F$ and $2k_\pm$ processes, see Fig.~\ref{fig:1}.
We find (using the gauge $\eta_{+} \eta_{-} = i$)
\bea
&& \left(
\begin{array}{c}
S^x\\
S^y\\
S^z
\end{array}
\right)_{2k_F} = -\frac{\cos[\sqrt{2\pi}\varphi_\rho + 2k_F x]}{\pi a_0} \times
\nonumber\\
&&\times
\left(
\begin{array}{c}
-\sin[\sqrt{2\pi}\theta_\sigma]\\
\cos[\gamma_F] \cos[\sqrt{2\pi}\theta_\sigma] +
\sin[\gamma_F] \cos[\sqrt{2\pi} \varphi_\sigma + \delta k_F x]\\
\sin[\sqrt{2\pi}\varphi_\sigma]
\end{array}
\right)
\nonumber\\
&& \to -\frac{\cos[\sqrt{2\pi}\varphi_\rho + 2k_F x]}{\pi a_0}
\left(
\begin{array}{c}
\pm 1\\
0\\
0
\end{array}
\right) .
\label{eq:8}
\eea
The last line of the above equation is somewhat symbolic, with zeros representing {\sl exponentially}
decaying correlations of the corresponding spin components, $S^{y,z}$.
Here
$\hat{z}$-component is disordered by strong quantum
fluctuations of {\sl dual} $\varphi_\sigma$ field,
as dictated by $[\varphi, \theta]$ commutation relation, see (\ref{eq:bos7}).
The $\hat{y}$-component does not order because
$\cos[\sqrt{2\pi}\theta_\sigma^{\text{cl}}]=0$.
Thus Cooper order found here in fact represents spin-density-wave
(SDW$_x$) order  at momentum $2k_F$
of the $\hat{x}$-component of spin density,
as discussed previously in Ref.~\onlinecite{sun07}.
Observe that $S^x$ ordering is of {\sl quasi-LRO} type
as it involves free charge boson, $\varphi_\rho$. As a result,
spin correlations do decay with time and
distance, but very slowly
$\langle S^x(x) S^x(0)\rangle \sim \cos[2k_F x]~x^{-K_\rho}$.

The result (\ref{eq:8}) also hints a possibility of truly long-range-ordered spin correlations
in the insulating state of the wire -- Heisenberg spin chain. There the charge field $\varphi_\rho$
is pinned by the relevant two-particle Umklapp scattering \cite{giamarchi-book}, which can
be mimicked by setting $K_\rho \to 0$ in the spin correlation function above. This is the
essence of the result to be discussed in Section~\ref{sec:heis} below.

Observe another interesting feature of Eq.~\ref{eq:8}: $S^y_{2k_F}$ has the appearance
of rotated by angle $\gamma_F$ component of the vector, whereas $S^x_{2k_F}$ and $S^z_{2k_F}$ remain unchanged. The question that arises is what  does $S^y_{2k_F}$ rotate into?
The full answer is provided
by considering $2k_F$-component  of the generalized \emph{helicity} operators
\bea
&&h^a_{2k_F}(x) = -(i/2k_F)\sum_{s,s'=\uparrow,\downarrow}\Psi^\dagger_s \sigma^a_{s,s'}
\hat{p}\Psi_{s'}|_{2k_F}\label{eq:staggered}\\
&=&  (i/2) \sum_{s=\uparrow,\downarrow} (R^\dagger_s \sigma^a_{s,s'} L_{s'} e^{-i2k_F x} -
L^\dagger_s \sigma^a_{s,s'} R_{s'} e^{i2k_F x}),\nonumber
\eea
where $a=\{0,x,y,z\}$. Note that
$h^0_{2k_F}(x)$ turns into well-known staggered dimerization operator $\epsilon(x)$
in the `spin chain limit' of the problem, when charge fluctuations disappear.
In terms of the right and left moving fermions the staggered ($2k_F$)  dimerization is given by
\bea
\epsilon= h^0_{2k_F} = (i/2) \sum_{s=\uparrow,\downarrow} (R^\dagger_s L_s e^{-i2k_F x} -
L^\dagger_s R_s e^{i2k_F x}).\label{eq:dimerization}
\eea
Its bosonized form, in terms of $\phi_\sigma$ and $\theta_\sigma$ fields (\ref{eq:bos5}),
\bea
h^0_{2k_F} &=& \frac{\cos[\sqrt{2\pi}\varphi_\rho + 2k_F x]}{\pi a_0}
\Big(\cos[\gamma_F] \cos[\sqrt{2\pi} \varphi_\sigma + \delta k_F x]\nonumber\\
&&- \sin[\gamma_F] \cos[\sqrt{2\pi}\theta_\sigma] \Big).\label{eq:bos_dimerization}
\eea
matches ``rotated" $S^y_{2k_F}$ in \eqref{eq:8} exactly.

Although similar looking, this operator is different from ``$2k_F$" component of
the density, described below in \eqref{eq:rho}. That one has charge boson $\varphi_\rho$
appearing under sine, see \eqref{eq:h^y} and \eqref{eq:tilde-rho},
while both $\epsilon$ and $\vec{S}$ fields are proportional
to the cosine of it (see also Ref.~\onlinecite{sfb}). The difference is important.
The $y-$component of spin and $h^0$ operators in the
original up and down spin basis can be written in a rather compact form,
\bea
S^y_{2k_F}= \cos[\gamma_F]\tilde{S}^y_{2k_F} +\sin[\gamma_F]\tilde{h}^0_{2k_F}\nonumber\\
h^0_{2k_F} =  \cos[\gamma_F]\tilde{h}^0_{2k_F} - \sin[\gamma_F]\tilde{S}^y_{2k_F},
\eea
where
\bea
\tilde{S}^y_{2k_F}(x)&=&\frac{1}{2}\sum_{\nu,\nu'=\mp} (R^\dagger_\nu \sigma^y_{\nu,\nu'}L_{\nu'}
e^{-i(k_\nu + k_{\nu'}) x} + \nonumber\\
&& + L^\dagger_\nu \sigma^y_{\nu,\nu'}R_{\nu'} e^{-i(k_\nu + k_{\nu'}) x})
\eea
and
\bea \tilde{h}^0_{2k_F}(x)
&=&\frac{i}{2} \sum_{\nu=\mp} (R^\dagger_\nu L_\nu e^{-i2k_\nu x}- L^\dagger_\nu R_\nu e^{i2k_\nu x}),
\eea
are, respectively,  the spin and $h^0$ operators in the $\mp$ basis.
Thus, in the limit of $\gamma_F \rightarrow \pi/2$, the $2k_F$ component of spin
along the $y-$direction in one basis appears as the $h^0$
operator in the second basis and vice versa.

Of the remaining staggered operators, $h^a$ ($a=x,y,z$) only $h^y$ is affected by rotation.
The $y$-component partially
``rotates" into the $2k_F$-part of the density operator,
$\tilde{\rho}_{2k_F}= \sum_{\nu=\mp} (R^\dagger_\nu L_\nu e^{-i2k_\nu x} + L^\dagger_\nu R_\nu e^{i 2k_\nu x} )$,
via the following relation
\bea
 h^y_{2k_F} = \cos[\gamma_F]  \tilde{h}^y_{2k_F} - \sin[\gamma_F] \tilde{\rho}_{2k_F}/2.
\eea
On the other hand the $2k_F$ component of density operator in the original spin basis,
$\rho_{2k_F}= \sum_{s=\uparrow,\downarrow} (R^\dagger_s L_s e^{-i2k_F x} + L^\dagger_s R_s e^{i2k_F x})$,
rotates into the y-component of the $h$-operator
\bea
\rho_{2k_F}/2 =\cos[\gamma_F] \tilde{\rho}_{2k_F}/2 + \sin[\gamma_F]  \tilde{h}^y_{2k_F} .
\label{eq:rho}
\eea
As before, tilde's are used to denote operators in the $\mp$ basis.
The bosonized forms for $\tilde{h}^y_{2k_F}$ and $\tilde{\rho}_{2k_F}$ are as follows:
\bea
\tilde{h}^y_{2k_F} = -\frac{ 1}{\pi a_0}\sin(\sqrt{2\pi}\phi_\rho +2k_Fx)\cos(\sqrt{2\pi}\theta_\sigma)
\label{eq:h^y}
\eea
and
\bea
\tilde{\rho}_{2k_F} = -\frac{ 2}{\pi a_0}\sin(\sqrt{2\pi}\phi_\rho +2k_Fx)\cos(\sqrt{2\pi}\phi_\sigma + \delta k_F x).
\label{eq:tilde-rho}
\eea
Note that  the charge content of  $h^y$ and density operator are the same
but the spin parts are different.

The following relation may be helpful in revealing the origins of $h^a$ and $\epsilon$ fields.
In the case of Heisenberg chain the staggered dimerization has meaning of the staggered
energy density, $\epsilon(x) = e^{i 2k_F x} \vec{S}(x) \cdot \vec{S}(x+a_0)$, where $a_0$ is
the lattice spacing. In the low-energy limit this expression turns into
$\epsilon(x) \propto(\vec{J}_R(x) + \vec{J}_L(x)) \cdot \vec{S}_{2k_F} (x')$, where the limit
$x'\to x$ must be taken. Short calculation shows that this leads to Eq.\eqref{eq:dimerization}
above. We now observe that, quite similarly to $\epsilon=h^0$,
the helicity operator $h^a$ (with vector index $a=x,y,z$) may be
understood as arising from the fusing of the spin current \eqref{eq:spin-currents}
with the $2k_F$-component  of the density field:
$h^a_{2k_F}(x) = (J_R^a(x) + J_L^a(x)) \rho_{2k_F} (x')$. Here again $x'\to x$ limit is understood.
One can check that all relations involving $h^a$ derived above follow from this observation.
In particular, we note that
\be
h^x_{2k_F} = \frac{2}{\pi a_0} \sin[\sqrt{2\pi}\varphi_\rho + 2k_F x] \sin[\sqrt{2\pi}\theta_\sigma] ,
\ee
implying that correlation function of this field decays with the same exponent ($K_\rho$)
as that of $S^x_{2k_F}$ discussed in the beginning of this Section.
It is useful to note that helicity disappears in the spin-chain limit of the problem,
together with the low-energy density fluctuations.

\section{Perturbative Approach}
\label{sec:pert}

The aim of this section is to show the limit
$\gamma_F \to 0$ can be obtained in a straightforward
perturbation expansion in $\alpha_R$. While results of this section parallel
conclusions of the previous {\sl two-subband} consideration in
Sections~\ref{sec:intersubband}, \ref{sec:bosonization}, the technical steps involved
are somewhat involved and are, in our opinion, of interest in its own right.
In addition, similar perturbative consideration of the {\sl impurity} effects later in
this work turn out to be very informative for understanding the physics. For these reasons
we choose to present the main steps of the perturbation theory in spin-orbit coupling $\alpha_R$.
The calculation starts very similar to \cite{gritsev05} but concludes with quite different steps.

The idea is to treat both magnetic field and Rashba terms as perturbations to the
standard single-channel ballistic quantum wire charge and spin sectors of which are
described by the decoupled Tomonaga-Luttinger Hamiltonians (\ref{eq:H-rho}) and (\ref{eq:H-sigma-0}).
The parameters $K_{\rho/\sigma}, u_{\rho/\sigma}$ of these unperturbed harmonic sectors
are given by eq.(\ref{eq:LLparameters}) but {\sl with} $\gamma_F = 0$. Spin backscattering
term (\ref{eq:H-sigma-B}) is in principle present (again with $\gamma_F = 0$)
but will not be required in the subsequent calculation.

Thus the perturbing terms are, see (\ref{eq:H0}), the Zeeman term,
\be
\hat{H}_{Z} =  - \Delta_z \int dx ~\Psi_s^\dagger
\frac{\sigma^z_{s s'}}{2} \Psi_{s'}
\label{eq:HZ1}
\ee
and the spin-orbit term given by,
\be
\hat{H}_{R} = \alpha_R \int dx~
\Psi^\dagger_s(x)\sigma^x_{s s'}(- i \frac{\partial}{\partial x}) \Psi_{s'}(x)
\label{eq:HR1}
\ee
where $\Psi_{s=\uparrow,\downarrow}(x)$ right and left movers of unperturbed single-channel
quantum wire
\be
\Psi_s = R_s e^{i k_F x} + L_s e^{-i k_F x}.
\label{eq:psi}
\ee
Bosonized expressions for $R/L$ operators parallels that in (\ref{eq:bos1},\ref{eq:bos2},\ref{eq:bos3})
where the subband index $\nu=\pm$ should be replaced by {\sl spin index} $s =\uparrow,\downarrow$.
In terms of charge and spin modes introduced in (\ref{eq:bos5}), dual pair $\varphi_s, \theta_s$ for a fermion
of a given spin projection $s$ is expressed as
\be
\varphi_s = (\varphi_\rho + s \varphi_\sigma)/\sqrt{2}~ , ~\theta_s = (\theta_\rho + s \theta_\sigma)/\sqrt{2},
\label{eq:bos6}
\ee
where the following correspondence for the right-hand-side of the equations
is understood: $s = \uparrow = +1$ and $s=\downarrow = -1$. It then follows that
\be
[\varphi_\lambda(x), \theta_{\lambda'}(x')] = \frac{i}{2} (1 - \text{sign}(x-x')),
\label{eq:bos7}
\ee
where $\lambda=s=\uparrow, \downarrow ~\text{or}~\rho,\sigma$.

We then find
\be
\hat{H}_{Z} = -\Delta_z \int dx ~(J^z_R + J^z_L) =
-  \frac{\Delta_z}{\sqrt{2\pi}} \int dx ~\partial_x \varphi_\sigma
\label{eq:HZ2}
\ee
and
\bea
&&\hat{H}_{R} = 2\alpha_R k_F \int dx~( J_R^x(x) -J_L^x(x)) \nonumber\\
&&
=\frac{2 \alpha_R k_F \eta_\uparrow\eta_\downarrow}{\pi a_0}\int dx \cos[\sqrt{2\pi} \varphi_\sigma] \sin[\sqrt{2\pi}\theta_\sigma],
\label{eq:HR2}
\end{eqnarray}
where $J_R^a$ and $J_L^a$ are the $a$-th components of chiral (right and left) spin-currents ($a=x,y,z$) defined as
\be
J_R^a = R^\dagger_s(x)\frac{\sigma^a_{s s'}}{2}R_{s'}(x)~,~
J_L^a = L^\dagger_s(x)\frac{\sigma^a_{s s'}}{2}L_{s'}(x) .
\label{eq:currents}
\ee
Note that $\hat{H}_{R}$, being determined by the {\sl difference} of right and left spin currents,
is odd under spatial inversion ${\cal P}$ which interchanges right and left movers.

Finally, we rescale fields $\theta_\sigma \to \theta_\sigma/\sqrt{K_\sigma}$ and
$\varphi_\sigma \to \sqrt{K_\sigma}\varphi_\sigma$
and account for the Zeeman term (\ref{eq:HZ2}) by a position-dependent {\sl shift}
$\varphi_\sigma \to \varphi_\sigma + \sqrt{K_\sigma/(2\pi)} \Delta_z x/u_\sigma$
so that
\bea
\hat{H}_R &=& \tilde{g}_R  \int dx \cos[\sqrt{2\pi K_\sigma} \varphi_\sigma + q_0 x]
\sin[\sqrt{\frac{2\pi}{K_\sigma}}\theta_\sigma],~~~~
\label{eq:HR3}\\
&& \tilde{g}_R = \frac{2 \alpha_R k_F }{\pi a_0}  \eta_\uparrow\eta_\downarrow,
~q_0 = \frac{K_\sigma\Delta_z}{u_\sigma} .\nonumber
\eea
Upto a factor of $K_\sigma \approx 1$ which appears here due to the rescaling
of the bosonic fields above, $q_0$ matches with $\delta k_F$ in (\ref{eq:delta-k_F})
for $\alpha_R =0$.

These standard transformations leave (\ref{eq:HR3}) as the only perturbation, and correspond
to $\Delta_z \gg \alpha_R k_F$ limit of the theory.
Observe that charge sector of the theory, (\ref{eq:H-rho}) and (\ref{eq:LLparameters}) with $\gamma_F =0$,
decouples from the spin  sector and does not generate any new term.

The calculation proceeds by expanding partition function in powers of $g_R$
\be
Z =\int e^{-S_0}( 1 - \int d\tau\hat{H}_R + \frac{1}{2}\int d\tau
d\tau'\hat{H}_R^2 + ... ),
\label{eq:Z}
\ee
where the unperturbed action $S_0$ is particularly simple
\be
S_0 =  \int dx d\tau  ~ \frac{u_\sigma}{2} [(\partial_x \varphi_\sigma)^2 +
(\partial \theta_\sigma)^2] -i \partial_\tau \varphi_\sigma
\partial_x \theta_\sigma.
\label{eq:S0-1}
\ee
The back-scattering term $\cos(\sqrt{8\pi K_\sigma} \varphi_\sigma +
2 \delta k_F  x)$, see (\ref{eq:H-sigma-B}),  is ignored as it generates terms at higher
order in perturbation theory (by coupling with $\hat{H}^2_R$).

Further details of the calculation are summarized in Appendix~\ref{sec:ap-C}.
The result is (\ref{ap:S2}) from where we can read off the leading correction to the spin Hamiltonian
\be
H^{(2)}_\sigma = \Big(\frac{\alpha_R k_F}{K_\sigma \Delta_z}\Big)^2
\frac{U(2k_F)}{(\pi a_0)^2 K_\sigma}
 \int dx   \cos[ \sqrt{\frac{8\pi}{K_\sigma}}~\theta_\sigma].
 \label{eq:H2}
 \ee
This compares well with our previous result (\ref{eq:H-sigma-C}) (remember that we  rescaled $\theta$ as
$\theta \to \theta/\sqrt{K}$ above) in the limit $\gamma_F \ll 1$, where the described calculation is applicable.

It is worth pointing out that perturbatively generated Cooper term,  instead of being
proportional to $(\alpha_R k_F/ v_F)^2$ as one would na\"ively expect from
a straight forward perturbation expansion of (\ref{eq:HR3}),
acquires a non-trivial dependence on both the back-scattering
amplitude, $U(2k_F)/(2\pi v_F)$, (via $K-1/K$ combination in (\ref{ap:f}))
and the Zeeman term $\Delta_z$ (via $q_0$ dependence of (\ref{ap:f})) and  is
proportional to $(\alpha_R k_F/ \Delta_z)^2 U(2k_F)/ v_F$.
A second thing to notice is  the crucial role of
Klein-terms (see (\ref{ap:S2})) in generating  the correct (positive) sign in eq.(\ref{eq:H2}).

\subsection{Arbitrary angle between SO and magnetic field}
\label{sec:tilt}

\begin{figure}[ht]
  \centering
   \includegraphics[width=2.0in]{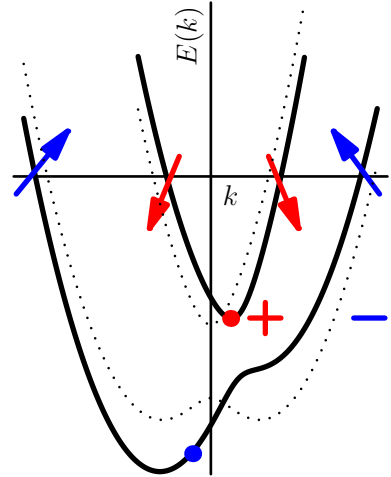}
       \caption{Occupied subbands $\epsilon_\pm$ for the case of
       non-orthogonal spin-orbital and magnetic field axes.
       Arrows illustrate spin polarization, as in Fig.~\ref{fig:1}.
    Filled dots indicate location of the center-of-mass for $(+)$ (red) and
    $(-)$ (blue) subbands. Dashed lines show $(\pm)$ subbands of Fig.~\ref{fig:1},
    corresponding to the orthogonal orientation, $\beta=0$, for comparison.}
  \label{fig:2}
\end{figure}

Perturbative approach is convenient for analyzing angular stability of the Cooper
phase. Let us consider the situation when the angle between the magnetic field
and spin-orbit directions is $\pi/2 - \beta$, so that $\beta=0$ corresponds to the
orthogonal orientation studied so far in this paper.
It is convenient to keep $\hat{z}$ as the direction of magnetic field, in which case
the spin-orbital term changes from $\alpha_R \hat{p} \sigma^x$ to
$\alpha_R \hat{p} (\sigma^x \cos[\beta] + \sigma^z \sin[\beta])$.
Following steps that led to (\ref{eq:HZ2},\ref{eq:HR2}) and (\ref{eq:HR3})
we obtain that $H_{R,\beta} = H_R^{(x)} + H_R^{(z)}$
\bea
H_R^{(x)} &=& \tilde{g}_R^{(x)}  \int dx \cos[\sqrt{2\pi K_\sigma} \varphi_\sigma + q_0 x]
\sin[\sqrt{\frac{2\pi}{K_\sigma}}\theta_\sigma] ,\nonumber\\
H_R^{(z)} &=&-\sqrt{\frac{2}{\pi}} \alpha_R k_F \sin[\beta] \int dx~\partial_x \theta_\sigma  ,
\label{eq:Hbeta1}
\eea
where first (second) equation represents contribution from $\sigma^x$ ($\sigma^z$) matrix.
The coupling constant of cosine term is only slightly modified in comparison with (\ref{eq:HR3}),
$ \tilde{g}_R^{(x)}  =  \tilde{g}_R \cos^2[\beta]$.
$H_R^{(z)}$ represents the main new feature of the non-orthogonal situation.
It is naturally absorbed into $H_R^{(x)}$ by a simple {\sl position-dependent} shift
of field $\theta_\sigma$ which results in
\bea
&&H_{R,\beta} = \tilde{g}_R^{(x)}  \int dx \cos[\sqrt{2\pi K_\sigma} \varphi_\sigma + q_0 x]  \nonumber\\
&& \times
\sin[\sqrt{\frac{2\pi}{K_\sigma}}\theta_\sigma + \delta x] ~;~
\delta = \frac{2 \alpha_R k_F \sin[\beta]}{\sqrt{K_\sigma} v_F}.
\label{eq:Hbeta2}
\eea
Note that both $\varphi$ and $\theta$ fields acquire oscillations now.
The consequence of this is that the generated Cooper term (compare with (\ref{eq:H2}))
\bea
H^{\text{C}}_\beta &=& \Big(\frac{\alpha_R k_F \cos[\beta] }{K_\sigma \Delta_z}\Big)^2
\frac{U(2k_F)}{(\pi a_0)^2 K_\sigma} \nonumber\\
&& \times
 \int dx   \cos[ \sqrt{\frac{8\pi}{K_\sigma}}~\theta_\sigma + 2 \delta x],
 \label{eq:Hbeta3}
\eea
{\sl does not} conserve momentum for $\beta \neq 0$. This important result is
 evident in the single particle spectrum, which
now reads ($\beta \neq 0$)
\be \epsilon_{\pm}  = \frac{k^2}{2m}  \pm \sqrt{(\alpha_R k)^2  +
(\Delta_z/2)^2    - \alpha k \Delta_z \sin\beta },
\label{eq:Hbeta4}
\ee
and is to be compared with (\ref{eq:epsilon_pm}), which describes $\beta =0$ situation.
The spectrum (\ref{eq:Hbeta4}) acquires an asymmetry
about the energy axis as seen in Fig.~\ref{fig:2}.
The top ($+$) subband shifts towards the right and the bottom ($-$) one
towards the left. As a result, the pair-tunneling does
not conserve momentum for $\beta \neq 0$,
as  is seen in (\ref{eq:Hbeta3}) above.

The momentum-mismatch $\delta$ de-stabilizes the Cooper order and eventually, for some
critical $\beta_c$, destroys it completely. The critical angle can be easily estimated by comparing
two spatial scales: $\xi_C$, which describes {\sl perfect} SDW order, and $1/\delta$,
which represents the scale on which momentum-nonconservation becomes pronounced.
Estimating $\xi_C \sim e^{\ell_c}$, where $\tilde{g}_R(\ell_c) \sim 1$, we obtain that
the Cooper order is destroyed when
\be
U(2k_F) (\alpha_R k_F \cos[\beta]/\Delta_z)^2 \approx \alpha_R k_F \sin[\beta].
\ee
Observe that this happens already for small angles, we can simplify the
expression for the critical angle
\be
\beta_c \propto \frac{U(2k_F)}{\Delta_z} \frac{\alpha_R k_F}{\Delta_z} \ll 1.
\label{eq:Hbeta6}
\ee
This angular sensitivity of the found Cooper order to mutual orientation of the magnetic
and spin-orbital directions can be used  as an experimental probe to
differentiate between it and other, spin-orbit independent, many-body instabilities
of interacting quantum wire.

\section{Isolated Impurity}
\label{sec:imp}

SDW$_x$ state manifests itself not only in spin correlations (\ref{eq:8}).
It turns out that its response to a weak potential scattering (impurity) is
rather non-trivial. We consider here a weak delta-function impurity $V(x) = V_0 \delta(x)$, with
strength $V_0$, located at the origin. The condition $V_0 \ll \Delta_c$ means
that impurity can be considered as a weak perturbation to the established
SDW phase. Since the latter is most robust in the limit of $\gamma_F \ll 1$,
this is the limit we consider in this Section. (The limit of strong impurity, $V_0 \gg \Delta_c$,
is rather standard: impurity destroys the SDW and the wire flows into an
insulator at low energies \cite{kane-fisher}.)

The interaction of electrons with an impurity potential $V(x)$ is
given by
\begin{eqnarray}
\hat{V} &=& \int dx ~V(x) \sum_{s=\uparrow,\downarrow } \Psi^\dagger_s(x) \Psi_s(x).
\end{eqnarray}
As usual, it is the {\sl backscattering} ($\rho_{2k_F}$) part of the density
that has to be considered, $\rho_{2k_F} = \sum_s (R^\dagger_s L_s + \text{h.c.})$.
Working in the two-subband basis of eigenstates $\nu=\pm$  of the
non-interacting Hamiltonian which includes Rashba spin-orbit term
and the Zeeman term, see (\ref{eq:3}), we derive for the $2k_F$-component
of the density $\rho(x)$ at the origin
\be
\rho_{2k_F} = \rho_{2k_F}^{\text{intra}} + \rho_{2k_F}^{\text{inter}},
\label{eq:rho-bs}
\ee
where the {\sl intra}-subband part is
\be
\rho_{2k_F}^{\text{intra}}  =  \cos[\gamma_F] (L^\dagger_+ R_+
 + L^\dagger_- R_- + \text{h.c.} ),
 \label{eq:rho-intra}
\ee
while the novel {\sl inter}-subband contribution is present due to the
non-orhogonality of spin states in $(\pm)$ subbands discussed in Section~\ref{sec:FreeElectrons},
 \be
\rho_{2k_F}^{\text{inter}} = \sin[\gamma_F] ( L^\dagger_- R_+
-L^\dagger_+ R_- + \text{h.c.}).
\label{eq:rho-inter}
\ee
Alternatively, one can think of these two contributions as representing
$\tilde{\rho}$ and $\tilde{h}^y$ contributions in \eqref{eq:rho}.
In terms of bosonic fields the density reads
\bea
\rho_{2k_F}(x=0)&=&\frac{-2  \sin[\sqrt{2\pi}\varphi_\rho]}{\pi a_0}\Big( \cos[\gamma_F]
\cos[\sqrt{2\pi}\varphi_\sigma]\nonumber\\
&& - \sin[\gamma_F] \cos[\sqrt{2\pi}\theta_\sigma]
\Big).
\label{eq:Vback}
\eea
We now observe that  setting $\theta_\sigma \to \theta_\sigma^{\text{cl}}= (m+\frac{1}{2}) \sqrt{\pi/2}$,
as appropriate for the Cooper phase (Section~\ref{sec:nature-C}),
{\sl nullifies} the backscattering component of the density (\ref{eq:Vback}).
The first (intra-subband) term gets killed by diverging fluctuations of
$\varphi_\sigma$, {\sl dual} to $\theta_\sigma$. Intriguingly, the second
(inter-subband) contribution is also zero because
$\cos[\sqrt{2\pi}\theta_\sigma^{\text{cl}}]=\pm \cos[\pi/2]=0$.

This argument can be made more precise by following calculations
described in Refs.~\onlinecite{review,orignac,egger}. We parametrize
$\theta_\sigma = \theta_\sigma^{\text{cl}} + \delta \theta$
and expand the relevant cosine term in Hamiltonian (\ref{eq:H-sigma-C})
to second order in fluctuations $\delta \theta$.
One obtains a {\sl massive} term $\propto \Delta_c (\delta \theta)^2$ in the Hamiltonian, which
causes exponential decay in correlation functions of the dual $\varphi_\sigma$ field.
In particular $\langle \cos \sqrt{2\pi}
\varphi_\sigma(0,\tau)\cos \sqrt{2\pi} \varphi_\sigma(0,\tau') \rangle$ will
decay as $\exp[-\Delta_c |\tau -\tau'| ]$. Thus  for an incoming particle
with energy $\omega \ll \Delta_c$, the inter-subband scattering
channel is absent.

Substituting $\theta_\sigma = \theta_\sigma^{\text{cl}} + \delta \theta$ in the second term in (\ref{eq:Vback})
converts it into $\gamma_F \sin[\sqrt{2\pi}\varphi_\rho]\sin[\sqrt{2\pi}\delta\theta]$.
Correlations of $\sin[\sqrt{2\pi}\delta\theta]$ are also
short-ranged
\bea
&&\langle\sin[\sqrt{2\pi}\delta\theta(\tau)]
\sin[\sqrt{2\pi}\delta\theta(\tau')] \rangle \approx
\frac{\Delta_c}{v_Fk_F}\times \nonumber\\
&&\sinh[K_0(\Delta_c |\tau-\tau'|)] ,
\eea
 where
$K_0(x)\sim e^{-x}/\sqrt{x}$ (for $x\gg 1 $) is the modified
Bessel function (see Ref.\onlinecite{review}).
The two exponentially decaying contibutions add up (in second order
perturbation theory in impurity strength $V_0$) to produce
an effective two-particle backscattering potential
$\propto (V_0^2/\Delta_c) \cos[\sqrt{8\pi}\varphi_\rho(0,\tau)]$.
This generated {\sl two-particle} impurity backscattering term,
however, is relevant only for strongly repulsive
interactions, $K_\rho < 1/2$.
We are thus left with {\sl irrelevant} impurity potential for as long as
$V_0 \ll \Delta_c$ is justified and for not too strong repulsion, $1/2 \leq K_\rho \leq 1$:
SDW$_x$ state is not sensitive to weak disorder!

The situation is similar to that in recently
proposed edge states in quantum spin Hall system
\cite{kane05}. There, gapless spin-up and spin-down excitations
propagate in opposite directions along the edge, which forbids
single-particle backscattering. Interacting electrons, however,
can backscatter off the impurity in pairs \cite{xu06,wu06}.

Our conclusion $\rho_{2k_F}\to 0$, eq.\eqref{eq:Vback},
rests on somewhat technical condition
$\cos[\sqrt{2\pi}\theta_\sigma^{\text{cl}}]=0$ and deserves a better understanding.
This is provided by the {\sl perturbative} calculation in Appendix~\ref{sec:ap-imp}.
The idea of the calculation is similar to that in Section~\ref{sec:pert}:
treat both impurity and
{\sl spin-orbit} terms as perturbation and generate spin-orbit-related corrections
to backscattering potential (proportional to $\gamma_F$ in (\ref{eq:Vback}))
{\sl perturbatively}.
In this way we can be certain that {\sl all} symmetry-allowed contributions
are accounted for.
This argument also makes it clear that the generated terms, being produced
by the SOI, have to be {\sl odd} under spatial inversion ${\cal P}$.
It is this symmetry that guaranties that $\sin[\sqrt{2\pi}\theta_\sigma]$ can not
be generated in the process.
This instructive calculation, carried in Appendix~\ref{sec:ap-imp}, indeed supports the
conclusions of the more formal {\sl two-subband} approach described in this Section.

We conclude this Section by pointing out experimental consequences of our
findings. Suppression of single-particle backscattering off weak impurity
under the outlined conditions implies an unusual {\sl negative} magnetoresistance
in one-dimensional wire. Indeed, the conductance of the wire with such
an impurity should remain at the perfect $G_{\rm wire} = 2e^2/h$ value for as long as
applied magnetic field is
directed perpendicular to the spin-orbital ($\sigma^x$ here) axis.
By either turning the  magnetic field off,
or simply rotating the sample (so that momentum mismatch between the two
subbands destabilizes SDW$_x$ order),
one should observe that conductance plateau deteriorates.
It will be destroyed completely in the zero-temperature limit.

\section{Heisenberg antiferromagnet with  DM
interaction and magnetic field}
\label{sec:heis}

In this section we will consider a closely related problem:
the effect of uniform and staggered  Dzyaloshisnkii-Moriya (DM) interaction on a spin
$S=1/2$ Heisenberg antiferromagnet (HAFM)  in the presence of a
magnetic field perpendicular to the DM vector.  We
start by demonstrating the equivalence between the quantum wire problem,
analyzed in the previous Sections, with that of Heisenberg spin chain
subject to an additional uniform DM interaction.
We then present a novel {\em chiral rotation} argument to show
that results of the two-subband approach in Section~\ref{sec:intersubband}
can be obtained from a straightforward combination of two
independent rotations for right- and left-movers.
We also discuss the difference between the {\em uniform} and
the {\em staggered} DM interaction, analyzed previously in Ref.~\onlinecite{ao}.

\subsection{Weak uniform DM interaction and strong Magnetic  Field}
\label{sec:uniformDM}

The Hamiltonian of an isotropic HAFM spin chain  is given by
\be
H_{\text{heis}} = J \sum_j
[S^{x}_{j}S^{x}_{j+1}+S^{y}_{j}S^{y}_{j+1}+S^{z}_{j}S^{z}_{j+1}].
\label{eq:Hheis}
\ee
 In the continuum limit,  the above Hamiltonian is most
conveniently described in terms of the  SU$(2)_1$
Wess-Zumino-Novikov-Witten model the  basic ingredients of which
are the uniform $SU(2)$ left $\vec{J}_L(x)$ and right $\vec{J}_R(x)$ spin currents,
defined via fermions in (\ref{eq:currents}),
 and the staggered magnetization $\vec{N}(x)$ \cite{gnt-book}.
 Another important component is provided by {\sl staggered} energy density (dimerization),
see eq.\eqref{eq:dimerization}, but this does not
develop any order in the presence of external magnetic field.
These operators will be used to represent
the  continuum limit of the spin,
\be
\vec{S}_j\rightarrow a_0 \Big( \vec{J}_L(x) +
\vec{J}_R(x)  + (-1)^{x/a_0} \vec{N}(x) \Big),
\label{eq:spin}
\ee
where $a_0$  is the lattice spacing and continuous space variable is introduced via
$x=j a_0$.
Continuum limit of (\ref{eq:Hheis}) is $H_{\text{heis}} = H_0 + H_{\text{bs}}$, where
\bea
H_0 & = &\frac{2\pi v}{3} \int dx ~\vec{J}_R \cdot \vec{J}_R + \vec{J}_L \cdot \vec{J}_L
\label{eq:Hheis-currents}\\
&& = \frac{v}{2} \int dx ~(\partial_x \varphi)^2 + (\partial_x \theta)^2.
\label{eq:Hheis-abel}
\eea
The first line constitutes nonabelian spin current formulation of the problem, which is convenient
for analyzing spin-rotation invariant (SU(2)) problems, while the second makes connection
with familiar abelian bosonization result (\ref{eq:H-sigma-0}): identifications
$(\varphi_\sigma, \theta_\sigma) \to (\varphi, \theta)$,  $u_\sigma \to v=\pi J a_0/2$
and $k_F \to \pi/(2a_0)$ finalize the connection. %$K_\sigma \to 1$,
The marginal backscattering term accounts for residual interaction between
spin excitations,
\be
H_{\text{bs}} = -g_{\text{bs}} \int dx \vec{J}_R \cdot \vec{J}_L .\label{eq:back-scattering}
\ee
Its coupling constant $g_{\text{bs}} > 0$ is known from extensive numerical studies
of the Heisenberg chain \cite{eggert96} : $g_{\text{bs}} = 0.23\times 2\pi v$.
This term is responsible for the fact that initial value of the Luttinger parameter
$K_\sigma = 1 + g_{\text{bs}}/2\pi v$ is greater than $1$.

There are no low energy charge fluctuations
as the charge is ``locked" to the lattice by relevant Umklapp processes. This allows one
to replace cosine of charged boson $\varphi_\rho$ in (\ref{eq:8}) by its expectation value $\lambda$:
$ \lambda = \langle \cos[\sqrt{2\pi} \varphi_\rho] \rangle$. The first line of (\ref{eq:8}) then establishes
abelian representation of the staggered magnetization
\bea
&& \left(
\begin{array}{c}
N^x\\
N^y\\
N^z
\end{array}
\right) = \frac{\lambda (-1)^x}{\pi a_0}
\left(
\begin{array}{c}
-\sin[\sqrt{2\pi}\theta]\\
\cos[\sqrt{2\pi}\theta]\\
-\sin[\sqrt{2\pi}\varphi]
\end{array}
\right).
\label{eq:stag-N}
\eea
Magnetic field is introduced via Zeeman term
\be
H_Z^{(z)} = - \sum_j g\mu_B B S^z_j \to -\Delta_z \int dx (J^z_R + J^z_L) ,
\label{eq:HZ4}
\ee
which is identical to (\ref{eq:HZ2}). The small upper index ($z$ here) indicates
the axis in spin space.
Finally, the Dzyaloshinskii-Moriya Hamiltonian describes asymmetric and odd under
spatial inversion interaction between spins
\be
H_{\text{DM}}^{(x)} = \sum_j \vec{D}\cdot \vec{S}_j \times \vec{S}_{j+1}.
\label{eq:DM}
\ee
Vector $\vec{D} = D \hat{x}$ fixes direction of spin anisotropy, which we choose to be along
spin-$\hat{x}$ direction, similar to the spin-orbit direction choice in (\ref{eq:H0}).

Using (\ref{eq:spin}) we see that continuum limit of the uniform DM Hamiltonian requires
the knowledge of the following operators
\bea
O_J(x,x') &=& J^y(x) J^z(x') - J^z(x) J^y(x') , \nonumber\\
O_N(x,x') &=& N^y(x) N^z(x') - N^z(x) N^y(x'),
\label{eq:OO}
\eea
where $x' = x+a_0$. The first of these follows from the well-known OPE for spin currents,
see for example eq.(25,26) of Ref.~\onlinecite{sfb} and set $\barz - \barz' = i a_0$:
\be
O_J(x,x') = \frac{1}{\pi a_0} (J^x_R(x) - J^x_L(x)) .
\ee
The second requires more work and is calculated using the definition (\ref{eq:stag-N}) and
bosonic OPEs (\ref{ap:ope1})
\be
O_N(x,x') = \frac{2 \lambda^2}{\pi a_0} (J^x_R(x) - J^x_L(x)) .
\ee
As a result
\be
H_{\text{DM}}^{(x)} = \tilde{D} \int dx ~(J^x_R - J^x_L) ~, ~\tilde{D} = \frac{D a_0}{\pi} (1 + 2 \lambda^2).
\label{eq:DM2}
\ee
Comparing (\ref{eq:DM2}) with (\ref{eq:HR2}) and (\ref{eq:HZ4}) with (\ref{eq:HZ2}) we observe
that the problem at hand is identical to that analyzed in Section~\ref{sec:pert}.
We immediately conclude that  the  $2k_F = \pi$  (staggered) components of spins
acquire long range order along the direction of $\vec{D}$
as soon as magnetic field is turned on along the orthogonal direction.
\be
\langle N^x \rangle = {\text{const}} ~(-1)^x,  ~\langle N^{y,z} \rangle = 0 .
\ee
It is useful to compare this equation with \eqref{eq:8}.
Note that the effect is controlled by the backscattering amplitude
$g_{\text{bs}} = 2(K_\sigma -1)$.

Except for the brief remark in section VIII of Ref.\onlinecite{bocquet},
the possibility of the long-range order in the magnetized spin chain
with asymmetric uniform DM interaction has not been
discussed in the literature, to the best of our knowledge.
We also would like to note technical similarities of our problem
with that of an anisotropic Heisenberg chain in a transverse field,
considered in Ref.\onlinecite{caux03}.

Experimentally, the field-induced long-ranged magnetic order
would be probably easiest to observe via thermodynamic measurements.
Similar to the case of copper benzoate \cite{dender97}, the order
should show up via an exponential suppression of the specific heat.
The magnitude of this suppression, which is controlled by the energy
gap $\Delta_c$ \eqref{eq:Delta-c}, should be a sensitive function of the magnetic field
orientation, reaching maximum when the field is orthogonal to the DM
axis as discussed above.

\subsection{Chiral rotation and tilted magnetic field}
\label{sec:rotation}
We now present an elegant argument, borrowed from the technically
closely related study of quantum kagom\'e antiferromagnet \cite{kagome},
exposing the nature of the Cooper
order to the fullest. Consider the situation, discussed in Section~\ref{sec:tilt},
of the {\em tilted} magnetic field
making an angle $\pi/2 -\beta$ with the DM axis direction (which
we keep at $\hat{x}$). The Hamiltonian describing this arrangement
is given by $H = H_0 + H_{\text{bs}} + V$, where
 $V$ includes Zeeman and DM fields
\be
V = \int dx \Big( d_R J^x_R - d_L J^x_L - h_1 (J^z_R + J^z_L)\Big).\label{eq:Zeeman-DM}
\ee
The parameters are
\be
d_R = \tilde{D} - h_2, ~d_L = \tilde{D} + h_2, ~h_1 = h \cos\beta, ~h_2 = h\sin\beta.
\label{eq:rot00}
\ee
We now rotate right (left) spin currents $\vec{J}_R$ ($\vec{J}_L$) about
$\hat{y}$ axis by angle $\theta_R$ ($\theta_L$) such that after the rotation
the $V$-term becomes
\be
V = - \int dx \Big( \sqrt{d_R^2 + h_1^2} ~M^z_R + \sqrt{d_L^2 + h_1^2} ~M^z_L\Big).
\label{eq:rot0}
\ee
The relation between old ($\vec{J}$) and new ($\vec{M}$) currents is simple
\be
\vec{J}_R = {\cal R}(\theta_R) \vec{M}_R,
~~~\vec{J}_L = {\cal R}(\theta_L) \vec{M}_L,\label{eq:Rotation}
\ee
here ${\cal R}$ is the rotation matrix
\bea
{\cal R}(\theta) =
\left(
\begin{array}{ccc}
\cos[\theta] & 0 & -\sin[\theta]\\
0 & 1& 0\\
\sin[\theta] & 0 & \cos[\theta]\\
\end{array}
\right).
\label{eq:rot}
\eea
The rotation angles are given by $\tan[\theta_R] = d_R/h_1$ and $\tan[\theta_L] = - d_L/h_1$.
The key reason behind these rotations is the observation that
unperturbed Hamiltonian $H_0$, being the sum of commuting right and left terms,
is invariant under the rotations. But the backscattering $H_{\text{bs}}$ is not, and
transforms into
\bea
H_{\text{bs}} &=& - g_{\text{bs}}\int dx \Big( \cos[\gamma] (M^x_R M^x_L + M^z_R M^z_L)
\nonumber\\
&&+ \sin[\gamma] (M^x_R M^z_L - M^z_R M^x_L) + M^y_R M^y_L\Big),
\label{eq:rot1}
\eea
where $\gamma=\theta_R - \theta_L$. The reason for this transformation
is of course that right and left currents are rotated in opposite directions
(and by different amounts). Observe that \eqref{eq:rot1} matches the interaction part
of \eqref{eq:H-sigma-current2}.

The nice thing about the rotation is that now both SO and Zeeman fields
can be taken into account by simple linear transformations of spin bosons.
Indeed, using abelian bosonization we observe that
\be
V=-\int dx \Big( \frac{v t_\varphi}{\sqrt{2\pi}} \partial_x \varphi_\sigma +
\frac{v t_\theta}{\sqrt{2\pi}} \partial_x\theta_\sigma\Big),
\label{eq:rot2}
\ee
where
\bea
&&t_\varphi = (\sqrt{d_L^2 + h_1^2} +  \sqrt{d_R^2 + h_1^2})/2v,\nonumber\\
&& t_\theta = (\sqrt{d_L^2 + h_1^2} -  \sqrt{d_R^2 + h_1^2})/2v. \label{eq:varphi-theta}
\eea
These linear terms are removed by shifts
\bea
&&\varphi_\sigma \to \varphi_\sigma + \frac{t_\varphi x}{\sqrt{2\pi}},\nonumber\\
&&\theta_\sigma \to \theta_\sigma + \frac{t_\theta x}{\sqrt{2\pi}} .
\eea
The price is that transverse components $M^{x,y}_{R/L}$ acquire
oscillating position-dependent factors
\be
M^+_R \to M^+_R e^{-i(t_\varphi - t_\theta) x} ,
~~~M^+_L \to M^+_L e^{i(t_\varphi + t_\theta) x}.
\label{eq:rot3}
\ee
The major consequence of this is that every term in $H_{\text{bs}}$,
with a single exception of $M^z_R M^z_L$ one,
 picks up oscillating factor
\bea
&&H_{\text{bs}} = - g_{\text{bs}} \int dx \Big( \cos[\gamma] M^z_R M^z_L + \nonumber\\
&& +\frac{\cos[\gamma] - 1}{4} (M^+_R M^+_L e^{i2t_\theta x} + {\text{h.c.}}) \nonumber\\
&& + \frac{\cos[\gamma] + 1}{4} (M^+_R M^-_L e^{- i 2t_\varphi x} + {\text{h.c.}}) \\
&& + \frac{\sin[\gamma]}{2} (M^z_L M^+_R e^{-i(t_\varphi - t_\theta) x} - M^z_R M^+_L
e^{i(t_\varphi + t_\theta) x} + {\text{h.c.}}) \Big). \nonumber
\label{eq:rot4}
\eea
This equation contains {\em all} arrangements that we have discussed in this paper.
Setting $t_\theta = 0$ corresponds to {\em orthogonal} ($\beta=0$)
orientation of spin-orbit (DM) and magnetic field axis. In that limit we
recover equations \eqref{eq:H-sigma-current1} ($\vec{J}$ there
corresponds to $\vec{M}$ here). Allowing for $\beta > 0$ leads us to
Section~\ref{sec:tilt}, results of which represent abelian version of the discussion
here. As shown there, the SDW$_x$ phase is stable in a finite angular interval
near $\beta = 0$.

It is worth noting that \eqref{eq:rot0} and \eqref{eq:rot2} imply
finite expectation values of $M^z_{R/L}$ currents:
\be
\langle M^z_{R/L}\rangle = \sqrt{d_{R/L}^2 + h_1^2}/(4\pi v) .
\ee
By virtue of \eqref{eq:rot} this implies finite expectation values of the uniform magnetization
along $z$ and $x$ axis, $J^{z/x}_R + J^{z/x}_L$, and uniform spin current,
$J^{z/x}_R - J^{z/x}_L$, along these two axes. That is,
\bea
\langle J^x_R - J^x_L \rangle = \sin[\gamma] \frac{t_\phi}{2\pi},
\langle   J^z_R - J^z_L \rangle = -\cos[\gamma] \frac{t_\theta}{2\pi},\nonumber\\
\langle  J^x_R + J^x_L \rangle = -\sin[\gamma] \frac{t_\theta}{2\pi},
\langle  J^z_R + J^z_L \rangle = \cos[\gamma] \frac{t_\phi}{2\pi}.
\eea
Equilibrium values of spin current and magnetization along  the $y-$direction vanish.
These relations complement discussion of the relations between staggered ($2k_F$)
components of various fields in Section~\ref{sec:nature-C}.

\subsection{Strong uniform DM interaction and weak Magnetic Field}
\label{sec:strongDM}

It is instructive to consider another `tricky' limit of the problem:
 strong uniform DM  interaction
 and a weak magnetic field, $D \gg \Delta_z$. The idea here is to account
 for the DM term {\sl exactly} on the lattice level, see (\ref{eq:strongDM1}),
 and treat the Zeeman term as a small perturbation to the obtained Hamiltonian
 (\ref{eq:strongDM3}).

The two axis, DM and Zeeman, are still orthogonal but it is more convenient now to
let $\vec{D}$ point along the $\hat{z}$-axis in spin space, while magnetic
field is pointing along $\hat{x}$. Following Ref.~\onlinecite{bocquet} we perform
unitary transformation of the {\sl lattice} Hamiltonian to absorb
the uniform DM term exactly,
\begin{eqnarray}
S_j^{+}\rightarrow  \tilde{S}_j^{+}
e^{i\alpha j}~,~ S_j^{-}\rightarrow \tilde{S}_j^{-}
e^{-i\alpha j} ~\text{and}~S^z_j \rightarrow  \tilde{S}^z_j,
\label{eq:strongDM1}
\end{eqnarray}
where  $\alpha = \arctan(D/J) \approx D/J$ for $D \ll J$.
The transformed Hamiltonian  is that of the XXZ spin chain with weak anisotropy
\be
\tilde{H} = \tilde{ J} \sum_j
[\tilde{S}^{x}_{j}\tilde{S}^{x}_{j+1}+\tilde{S}^{y}_{j}\tilde{S}^{y}_{j+1}+
\frac{J}{\tilde{J}}\tilde{S}^{z}_{j}\tilde{S}^{z}_{j+1}] ,
\label{eq:strongDM2}
\ee
where $\tilde{J} = \sqrt{J^2 + D^2}$. Magnetic field term, however,
acquires position dependence under this transformation
\be
\tilde{H}_Z^{(x)} = -\frac{\Delta_z}{2}\sum_j (\tilde{S}^{+}_j e^{i \alpha j} + \tilde{S}^{-}_j e^{-i\alpha j}).
\label{eq:strongDM3}
\ee
Bosonizing it according to the rules described in Sec.~\ref{sec:bosonization} and \ref{sec:pert},
we obtain
\begin{eqnarray}
\tilde{H}_Z^{(x)} = \frac{\eta_\uparrow \eta_\downarrow \Delta_z}{\pi a_0}
 \int dx \sin[\sqrt{2\pi K} \varphi] \cos[\sqrt{\frac{2\pi}{K}} \theta  + \alpha x] ,
\label{eq:strongDM4}
\end{eqnarray}
where Luttinger parameter of the XXZ chain (\ref{eq:strongDM2}) is given by
$K^{-1} = 1 - \arccos[J/\sqrt{J^2 + D^2}]/\pi \approx 1 - D/(\pi J)$.
The shift of $\theta$ field by $\alpha x$ can be easily understood from equation (\ref{eq:DM2}),
adapted for the DM axis along $\hat{z}$-direction in spin space:
$H_{\text{DM}}^{(z)} = \tilde{D} \int dx (J^z_R - J^z_L) \propto \tilde{D} \int dx ~\partial_x \theta$
which indeed can be accounted for by shifting $\theta$.

Being of nonzero conformal spin, the Zeeman term \eqref{eq:strongDM4}
generates under RG
two new terms of importance to us:
$\tilde{H}_\varphi \propto (\Delta_z^2 D/J) \cos[\sqrt{8\pi K} \varphi]$,
of scaling dimension $2K = 2 + 2D/(\pi J) > 2$,  and
$\tilde{H}_\theta = G_\theta \int dx \cos[\sqrt{\frac{8\pi}{K}} \theta  + 2\alpha x]$
with scaling dimension $2/K = 2 - 2D/(\pi J) < 2$. The position-dependent phase
$2\alpha x$, nonetheless,
makes it {\sl irrelevant}.

Indeed, the coupling constant can be estimated, following momentum-shell RG
of Appendix~\ref{sec:ap-shell}, as $G_\theta \propto (\Delta_z/v)^2 (K-1/K)^2 \approx (\Delta_z D/J)^2$,
see (\ref{ap:2order-bs}) and (\ref{ap:Bbs}).
Neglecting the oscillating phase for the moment, we can easily estimate the correlation
length corresponding to the Cooper order as $\xi_c = \exp[\ell_c]$, where
$G_\theta(\ell_c) = G_\theta(0) \exp[(2 - 2/K) \ell_c] \approx 1$, so that
$\ell_c = (\pi J/D) \ln[J/(\Delta_z D)]$.

This length is to be compared with $\xi_\alpha = 1/\alpha \approx J/D$, which is the length
on which the cosine in  $\tilde{H}_\theta$ changes {\sl sign}.  We observe that
\be
\ln[\xi_c] = \frac{\pi J}{D} \ln\Big(\frac{J}{\Delta_z D}\Big) \gg \ln[\xi_\alpha] = \ln(\frac{J}{D})
\ee
which implies that fast oscillations make it impossible
for the $\tilde{H}_\theta$ to reach strong coupling. Hence no order can come from this term.

The other term, $\tilde{H}_\varphi$, {\em appears} to be irrelevant and it is tempting to
disregard it altogether. One however must be careful and recall discussion in Section~\ref{sec:rg}.
Particularly, observe that $\tilde{H}_\varphi$ may fall in region {\bf II} (with $y_\sigma(0) < 0$).
Whether or not this happens is in principle a subject of precise calculation which however can be
avoided here. Indeed analysis of Section~\ref{sec:rotation}, taken together with
results of Sec.~\ref{sec:rg}, shows that Cooper instability develops for
arbitrary $\gamma$ as long as $t_\theta=0$.

We thus conclude that $\tilde{H}_\varphi$ {\em must} belong to region {\bf II}
in Figure~\ref{fig:rgflow}, even though this conclusion is not
at all obvious from the abelian bosonization analysis sketched above.
We chose to present this subtle point in order to demonstrate the
power of non-abelian current formulation in Section~\ref{sec:rotation}.

Note finally that $\tilde{H}_\varphi$ still describes Cooper order even though
it is written in terms of $\varphi$ field. The reason for this illustrates
another shortcoming of abelian formulation. Unitary transformation \eqref{eq:strongDM1}
is the lattice version of rotation of right (left) currents by $\theta_R = \pi/2$ ($\theta_L = -\pi/2$)
in the previous Section. (In this limit $\Delta_z=0$ and magnetic field is to be added later as a perturbation.)
Such a rotation corresponds to $\gamma=\pi$ which replaces standard backscattering
$(M^+_R M^-_L  + {\text{h.c.}})$ with $(M^+_R M^+_L + {\text{h.c.}})$. In terms of
abelian bosonization this corresponds to interchange
$\cos\sqrt{8\pi} \varphi \leftrightarrow \cos\sqrt{8\pi}\theta$. Moreover, this same notation
changes the sign of $M^z_R M^z_L$ term, which corresponds to interchange
$K \leftrightarrow 1/K$, which completes the duality mapping.
Observe again the ease with which current formulation of Section~\ref{sec:rotation}
leads to conclusion.

\begin{figure}[ht]
  \centering
   \includegraphics[width=2.0in]{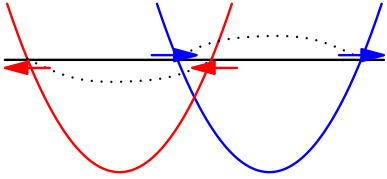}
       \caption{Band structure in the absence of magnetic field.
       Note the absence of the gap between the subbands at $k=0$.}
  \label{fig:nomag}
\end{figure}

In simpler terms, accounting for DM (or, equivalently, spin-orbit) term only
amounts to working in the basis of eigenstates of operator $\sigma_x$.
These eigenstates are described simply by two {\em horizontally} shifted parabolas
in Figure~\ref{fig:nomag}. Denoting the right (left) subbands by index $1$
($|1\rangle = |\rightarrow\rangle$) and $2$
($|2\rangle = |\leftarrow\rangle$) we immediately
observe that Cooper scattering discussed previously corresponds to the process
like $R^\dagger_1 L_1 L^\dagger_2 R_2$. (So that initial pair $L_1 R_2$
scatters to final $R_1 L_2$ state. Note that the other possibility, $L_1 \to L_2$
and $R_2 \to R_1$ is forbidden by orthogonality of single-particle basis states,
$\langle 1|2\rangle = \langle \rightarrow| \leftarrow\rangle =0$.) Bosonizing this in a standard
way we find that
(recall $R^\dagger_1 L_1 \sim \exp[-i\sqrt{4\pi} (\phi_{1R} + \phi_{1L})]$)
$R^\dagger_1 L_1 L^\dagger_2 R_2 \sim \exp[-i\sqrt{4\pi}(\varphi_1 - \varphi_2)]
= \exp[-i\sqrt{8\pi}\varphi_\sigma]$. Thus indeed the Cooper process is
written in terms of $\varphi_\sigma$ field in the new basis.

Observe that this conclusion is special to the limit of no magnetic field, $\Delta_z =0$.
As soon as the field is on, $\Delta_z \neq0$, the band structure changes and
horizontally shifted subbands $(1, 2)$ turn into vertically shifted pair $(-,+)$
used in this paper (Fig.~\ref{fig:1}): arbitrary small $B$ leads to the gap $\Delta_z$ at the
point $k=0$, see \eqref{eq:epsilon_pm}. We thus conclude once again that
$\Delta_z =0$ is a rather singular limit, not much suited for the consideration of
general situation with both spin-orbit and Zeeman fields present.

\subsection{HAFM chain with staggered DM interaction and magnetic field}
\label{sec:stagDM}

For the sake of completeness and uniformity of the presentation, we
re-visit here the well-studied case of interplay between {\sl staggered} DM
interaction and Zeeman magnetic field, described in Refs.~\onlinecite{ao}
and \onlinecite{oshikawaJPSJ}.

Staggered DM interaction along $\hat{x}$-axis is described by
\be
H_{\text{sDM}}^{(x)} = D \sum_j (-1)^j (S^y_j S^z_{j+1} - S^z_j S^y_{j+1}).
\label{eq:sDM}
\ee
Its continuum limit requires the knowledge of
\bea
O_{JN}(x,x') &=& N^y(x) J^z(x') + J^z(x) N^y(x')\nonumber\\
&& - J^y(x) N^z(x') - N^z(x) J^y(x'),
\label{eq:O-JN}
\eea
where, as before in (\ref{eq:OO}), we set $x' = x+a_0$.
It is an easy calculation, using eq.26 of Ref.\onlinecite{sfb},
to find that in the absence of magnetic field
$O_{JN}=0$ identically. The situation changes when magnetic field (along $\hat{z}$-axis)
 is present.
Using bosonization, magnetic field is accounted for (see (\ref{eq:HR3}) and line above it)
by the shift $\sqrt{2\pi K} \varphi \to \sqrt{2\pi K} \varphi + q_0 x$.
Moreover, the main effect of the field here is contained in $q_0 = \Delta_z/v$
and we can keep $K =1$ in all calculations below.
Observing that a shift in $\varphi$ implies equal shifts in $\phi_{R/L}$ fields
(so that $\theta$ does not change) and working backward through
(\ref{eq:8}), (\ref{eq:currents}) and (\ref{eq:bos1}) we conclude,
following Ref.\onlinecite{ao}, that
in the presence of the magnetic field, spin excitations {\sl along} the field ($\hat{z}$-axis here)
and {\sl transverse} to it ($\hat{x},\hat{y}$ axis) have minima at different momenta.
Namely, while $J^z$ is still centered at $q=0$, {\sl transverse} components
of spin current $J^{x,y}$ acquire minima
at $q=\pm q_0 = \pm 2\pi m$, where $m$ is the magnetization.
In addition, $N^z$ shifts from $\pi$ to $q=\pi \pm 2\pi m$ while
staggered transverse components $N^{x,y}$ remain at $2k_F = \pi$ point.
As a result, the product $N^y J^z$ (first line in (\ref{eq:O-JN})) retains
its zero-field structure and continues to remain at zero. At the same time
the other combination, $J^y N^z$ (second line in (\ref{eq:O-JN})),
splits into {\sl slow} ($e^{iq_0(x-x')}$) and {\sl fast} ($e^{iq_0(x+x')}$)
oscillating pieces which do not cancel each other anymore.
The remaining calculation is most conveniently performed using fermionic representations
of spin currents (\ref{eq:currents}) and longitudinal magnetization
$N^z = \frac{1}{2}[R^\dagger_\uparrow L_\uparrow - L^\dagger_\downarrow R_\downarrow] e^{-i q_0 x}
+ \text{h.c.}$. Fusing right (left) movers of like spin using (\ref{ap:GRL})
we obtain, for example,
\be
(J^+_R(x) + J^+_L(x)) N^z(x') = -\frac{\sin[q_0 (x-x')]}{2\pi (x-x')} (R^\dagger_\uparrow L_\downarrow
+ L^\dagger_\uparrow R_\downarrow).
\label{eq:J+N}
\ee
Bosonizing this expression (note that it is not sensitive to the {\sl sign}
of the coordinate difference $(x-x')$) we finally obtain
\be
O_{JN}(x,x') = - \frac{q_0 \lambda}{2\pi^2 a_0} \cos[\sqrt{2\pi} \theta(x)] .
\label{eq:O-JN2}
\ee
The same result can be obtained, after somewhat longer calculation, using bosonized forms of spin currents
and magnetization from the very beginning.
The continuum limit of staggered DM term then follows
\be
H_{\text{sDM}}^{(x)} = \frac{D q_0 \lambda}{\pi^2 a_0} \int dx ~\cos[\sqrt{2\pi} \theta(x)] .
\label{eq:sDM2}
\ee
This is a highly relevant operator (scaling dimension $1/2$),
 the coupling constant $G_{\text{s-dm}}$
of which grows as $G_{\text{s-dm}}(\ell) = G_{\text{s-dm}}(0) \exp[3\ell/2]$.
This, and the dependence of its initial value on the combination
$D \Delta_z$ (which enters via dependence on $q_0$),
leads to the energy gap in the system that scales as $(D \Delta_z)^{2/3}$, exactly as
Ref.\onlinecite{ao}
found originally. Note finally that (\ref{eq:sDM2}) and (\ref{eq:8}) implies that the spins order
(in a staggered way) along $\hat{y}$-axis, orthogonal to both DM and magnetic field directions.

Note that although we have treated staggered DM term (\ref{eq:sDM}) as a perturbation to the
spin chain subject to magnetic field (which comes in via the momentum $q_0$ here), we obtained
the same strongly relevant result (\ref{eq:sDM2}) as the authors of Ref.~\onlinecite{ao} did.
But Ref.~\onlinecite{ao} arrived at it from a different limit:
the authors used {\sl staggered} version of (\ref{eq:strongDM1}) to
account for DM piece (\ref{eq:sDM}) exactly and
then added magnetic field as a perturbation. The outcome (\ref{eq:sDM2}) is obtained in both cases
simply because it is more relevant (dimension $1/2$) than the magnetic field (Zeeman) term (dimension $1$):
no matter how small $G_{\text{s-dm}}$ initially is, it controls the physics. The $N^y$-order is present
for any ratio of $G_{\text{s-dm}}/\Delta_z$ ratio.

\section{Implications for ESR experiments}
\label{sec:esr}

In recent years, electron spin resonance (ESR) technique has become a leading
experimental candidate for probing anisotropic terms  in  the spin chain.
Recent theoretical impetus to this field has been provided in an important work
by Oshikawa and Affleck \cite{oa-esr,oa} who discussed  limitations of
earlier theoretical work \cite{kubo-tomita,mori-kawasaki1,mori-kawasaki2} and improved and extended upon them by  using powerful modern
theoretical techniques. The emphasis  in  \cite{oa} has been to study the role of anisotropic terms,
in particular the staggered DM interaction, see Section~\ref{sec:stagDM},
and exchange anisotropy terms, in modifying the resonance position and the line width.
However, ESR in a spin chain with uniform DM interaction was not considered.
We will follow a closely related work by De Martino {\sl et al.} \cite{egger02,egger04}, who
considered ESR for carbon nanotubes with spin-orbit coupling, to investigate the
modifications brought upon by the uniform DM term  on the ESR spectra.

We consider  Faraday configuration in which the static magnetic field and  oscillating field
are orthogonal to each other. The oscillating driving field (of frequency $\omega$)
is in the microwave frequency
regime and for all practical purposes the spatial modulation of the field can be ignored.
The ESR intensity is given by the transverse (to the static magnetic field)
spin structure factor at $q=0$ and frequency $\omega$,
\be
I(\omega)=  \int dt dx ~e^{i\omega t} \sum_{r,r'=R/L}  \langle J^{+}_r(x,t)J^{-}_{r'}(0,0)\rangle .
\label{esr:1}
\ee
As before, the Hamiltonian, $H= H_0 + H_{bs} +  V$, consists of the free part (\ref{eq:Hheis-currents}),
the back scattering part (\ref{eq:back-scattering}), and the DM and  Zeeman terms
contained in  the $V$ term (\ref{eq:Zeeman-DM}). The angle between DM and magnetic field
directions is $\pi/2 - \beta$, see Section~\ref{sec:rotation}.

Our aim here is to present the basic picture of ESR response for the case of the spin chain
with {\sl uniform} DM interaction (equivalently, quantum wire with spin-orbit interaction).
For this {\sl zeroth order} description we omit the backscattering interaction \eqref{eq:back-scattering}
between the spin currents altogether. Under this drastic approximation right $\vec{J}_R$ and
left $\vec{J}_L$  spin currents are decoupled.
This implies the intensity \eqref{esr:1} is the sum of right and left contributions,
$I(\omega)=I_R(\omega) + I_L(\omega)$.

To account for the simultaneous presence of DM and Zeeman terms, we now rotate
the right and left currents about the $y-$axis as described in Section~\ref{sec:rotation},
see in particular (\ref{eq:Rotation}) and (\ref{eq:rot}). As the backscattering \eqref{eq:rot1} is neglected,
the full Hamiltonian is given by
\bea
H &=&  \int dx \sum_{a=x,y,z} \frac{2\pi v}{3} \Big(M^a_RM^a_R + M^a_L M^a_L\Big)  \nonumber\\
&& -\lambda_R M^z_R -\lambda_L M^z_L ,
\label{eq:H-rot}
\eea
where $\lambda_{R/L}= \sqrt{d_{R/L}^2 + h_1^2} = v(t_\varphi \mp t_\theta)$.

We now focus on the contribution of the right spin currents,
\be
I_R(\omega)=  \int dt dx ~e^{i\omega t}  \langle J^{+}_R(x,t)J^{-}_{R}(0,0)\rangle ,
\label{esr:2}
\ee
which, in terms of  the rotated currents, reads
\bea
I_R(\omega) &=& \int dt dx ~e^{i\omega t}  \Big(\cos^2\theta_R \langle M^{x}_R(x,t) M^{x}_R(0)\rangle
\\
&&  +  \langle M^{y}_R(x,t)M^{y}_R(0)\rangle +
\sin^2\theta_R \langle M^{z}_R(x,t)M^{z}_R(0)\rangle  \Big) .\nonumber
\eea
The cross-terms of
the kind $\langle M^{x}_R(x,t)M^{z}_R(0)\rangle=0$ due to the absence of
coupling between $x$ and $z$ components in the Hamiltonian \eqref{eq:H-rot}.
Further, switching to $M^\pm_R$ combinations of the currents,
the intensity becomes
\bea
I_R(\omega) &=& \int dt dx ~e^{i\omega t} \Big(
\sin^2\theta_R \langle M^{z}_R(x,t)M^{z}_R(0)\rangle \nonumber\\
&& + \frac{(\cos^2\theta_R-1)}{4}
\langle M^{+}_R(x,t)M^{+}_R(0) + \text{h.c.}\rangle \\
&& +  \frac{(\cos^2\theta_R +1)}{4}\langle M^{+}_R(x,t)M^{-}_R(0) +  \text{h.c.}\rangle  \Big) .\nonumber
\label{esr:2}
\eea
The second line in this expression contributes zero
as it requires anomalous averages of the kind
$\langle \Psi_{R,\uparrow}(x,t)\Psi_{R,\uparrow}(0,0) \rangle$
which are absent in \eqref{eq:H-rot}. We now absorb ``right" magnetic field $\lambda_R$
via the shift of right boson so that $M^+_R \to M^+_R e^{-i \lambda_R x/v}$ 
(compare with \eqref{eq:rot3}).
It is worth noting that by rotating the spin currents we have mapped the problem
of the spin chain with uniform DM and magnetic fields to that of the chain in field $\lambda_R$
($\lambda_L$) for its right (left) moving components. This allows to borrow results of Ref.~\onlinecite{oa}
and conclude that the first line in \eqref{esr:2} does not contribute to $I_R$ while the last line gives
\be
I_R(\omega) = \frac{(\cos^2\theta_R +1)}{2} \omega \delta(\omega - |\lambda_R|) .
\label{esr:3}
\ee
Clearly the contribution of the left-moving sector is obtained by replacing $R \to L$ in the expression
above. Thus the ESR signal consists of {\em two} sharp lines, as previously discussed
for the case of carbon nanotube in
Refs.~\onlinecite{egger02,egger04},
\bea
I(\omega) &=& (\cos^2\theta_R +1) \lambda_R \delta(\omega - |\lambda_R|) \nonumber\\&&
+ (\cos^2\theta_L +1) \lambda_L \delta(\omega - |\lambda_L|) .
\label{esr:4}
\eea
The distance between the lines is $\lambda_L - \lambda_R = 2 v t_\theta$.
The relative strength of the two lines is
\be
\frac{I_R(\lambda_R)}{I_L(\lambda_L)} = \sqrt{\frac{d_L^2 + h_1^2}{d_R^2 + h_1^2}} .
\ee
The ratio is always $1$ for $\beta=0$ (orthogonal orientation of DM and magnetic field axes),
see \eqref{eq:rot00}. Note that this is exactly the configuration in which SDW order develops,
as described in detail in Section~\ref{sec:rg}. The ordering is driven by the Cooper term,
$H_\sigma^C$ in \eqref{eq:H-sigma-current1} (equivalently, the term proportional to
$(\cos[\gamma]-1)$ in \eqref{eq:rot4}), which comes from the backscattering process
as is made clear by the discussion in Section~\ref{sec:rotation}. The result \eqref{esr:4},
obtained by neglecting backscattering, is clearly not applicable at low temperature
where the SDW (Cooper) instability develops. Well below the ordering temperature
the system is described by the sine-Gordon model excitations of which are massive
kinks \cite{zvyagin05}. But even above the ordering temperature in the disordered
phase (which, for $\beta > 0$ extends all way down to zero
temperature, see Section~\ref{sec:tilt}) we expect the backscattering to affect
the ESR signal. Whether or not it would lead to the finite linewidth of the two lines
in \eqref{esr:4} we do not understand yet and leave this interesting question
(see Ref.~\onlinecite{oa} for detailed discussion of some technical subtleties)
for future studies.

\section{Conclusions}
\label{sec:conclusions}

Spin-orbital interactions result in reduction of spin-rotational symmetry
from $SU(2)$ to $U(1)$ in one-dimensional quantum wires
and spin chains. This reduction, however, is not sufficient to
change the critical (Luttinger liquid) nature of the one-dimensional
interacting fermions. The situation changes dramatically once
external magnetic field
is applied, as we have shown in this paper. Most interesting situation
occurs when the applied  field is oriented
along the axis orthogonal to the spin-orbital (or, Dzyaloshinskii-Moriya,
in case of spin chain) axis of the wire. The resulting combination
of two non-commuting perturbations, 
 taken together with
electron-electron interactions, leads to a novel spin-density-wave
order in the direction of the spin-orbital axis.

The physics of this order is elegantly described in terms of
spin-non-conserving (Cooper) pair tunneling processes between
Zeeman-split electron subbands.
The tunneling matrix element is finite only due to the presence of
the spin-orbit interaction, which allows for spin-up to spin-down (and vice versa)
conversion.

The resulting SDW state affects both spin and charge properties of the wire.
In particular, it suppresses effect of (weak) potential impurity, resulting
in the interesting phenomena of negative magnetoresistance in one-dimensional
setting, as described in Section~\ref{sec:imp}.

SDW ordering acquires true long-range nature in the case of spin chain,
where charge fluctuations are absent. The staggered moment  points
along the DM axis and is orthogonal to the applied magnetic field.

Even when the magnetic field and spin-orbital directions are not orthogonal,
an arrangement when the critical Luttinger state survives down to the lowest temperature
(see Section~\ref{sec:tilt}), the problem remains interesting. In this geometry
an ESR experiment should reveal two separate lines, which represent 
separate responses of right- and left-moving spin fluctuations in the system.

It is worth pointing out that unusual consequences of the interplay
of spin-orbit and electron interactions are not restricted to one-dimensional systems only.
We have recently shown \cite{vdW07} that Coulomb-coupled two-dimensional
quantum dots acquire a novel van der Waals-like anisotropic interaction
between spins of the localized electrons. The strength of this Ising interaction is determined
by the forth power of the Rashba coupling  $\alpha_R$.

We hope that our work will stimulate experimental search and studies of strongly
interacting quasi-one-dimensional systems with sizable spin-orbital interaction,
in particular regarding their response to the (both magnitude and direction)
applied magnetic field and/or magnetization. ESR studies of spin chain materials
with uniform DM interaction are very desirable as well.

We would like to thank I. Affleck, L. Balents, G. Fiete, C. Kane, D. Mattis, K. Matveev,
E. Mishchenko, J. Moore, L. Levitov, J. Orenstein, M. Oshikawa, M. Raikh, K. Samokhin,
and Y.-S. Wu for useful discussions and suggestions at various stages of this work.
O.S. thanks the Petroleum
Research Fund of the American Chemical Society for the financial support
of this research under the grant PRF 43219-AC10.

\appendix
\section{Derivation of intra-subband Hamiltonian (\ref{eq:Hintra})}
\label{sec:ap-intra}

Non-interacting (kinetic energy) part of the $\nu$-th subband ($\nu=\pm$) Hamiltonian
reads
\be
H^0_{\text{intra}} = \frac{v_F}{2} \int dx ~(\partial_x \phi_\nu)^2 + (\partial_x \theta_\nu)^2.
\label{ap:H0-intra}
\ee
The intra-subband interaction term is given by the part of (\ref{eq:Hint}) which involves
only densities from the $\nu$-th subband
\be
H'_{\text{intra}} =\frac{1}{2} \int dx dx' U(x-x') \rho_\nu(x) \rho_{\nu}(x'),
\label{ap:H-intra1}
\ee
where the density in the $\nu$-th subband is expressed with the help of (\ref{eq:3}) as
\bea
\rho_\nu &=& R^\dagger_\nu R_\nu + L^\dagger_\nu L_\nu + \cos[\gamma_\nu] \nonumber\\
&&\times \Big( e^{-i2k_\nu x} R^\dagger_\nu L_\nu
+ e^{i2k_\nu x} L^\dagger_\nu R_\nu \Big).
\label{ap:rho-nu}
\eea
Its bosonized form follows
\be
\rho_\nu = \frac{1}{\sqrt{\pi}} \partial_x \phi_\nu - \frac{\cos[\gamma_\nu]}{\pi a_0} \sin[\sqrt{4\pi} \phi_\nu + 2k_\nu x] ,
\label{ap:rho-nu-bos}
\ee
where the first (second) term represents uniform ($2k_\nu$) parts of density.
The interaction term (\ref{ap:H-intra1}) then naturally splits into a sum of two contributions
\bea
&&H'_{\text{intra}} = H'_{0} + H'_{2k_\nu} , \nonumber\\
&& H'_{0} =\frac{U(0)}{2\pi} \int dX ~(\partial_X \phi_\nu)^2 ,
 \label{ap:H'-0}\\
&& H'_{2k_\nu} =\frac{\cos^2[\gamma_\nu]}{2(\pi a_0)^2} \int dx dX ~U(x)
 \sin[\sqrt{4\pi} \phi_\nu(X+x/2) + \label{ap:H'-2kf}\nonumber\\
&&+2k_\nu (X+x/2)]
  ~\sin[\sqrt{4\pi} \phi_\nu(X-x/2) + \nonumber\\
  &&+2k_\nu (X-x/2)],
 \eea
where $x \to x-x'$ and $X=(x+x')/2$ are the relative and center-of-mass coordinates, respectively.

Next, following Ref.~\onlinecite{giam00},
we fuse the two sines in (\ref{ap:H'-2kf}) (the result is denoted as $S$ below)
using (\ref{eq:bos4}) and OPE identities (\ref{ap:ope1},\ref{ap:corr1}) to obtain
\bea
S(x,X) &=& \frac{a_0^2}{4 x^2} \sum_{\mu=\pm}  e^{i \mu 2k_\nu x} \times \label{ap:S}\\
&&\times \exp\{i\mu\sqrt{4\pi} [\phi(X+x/2) -\phi(X-x/2)]\} .\nonumber
\eea
Performing gradient expansion in $x$, neglecting boundary contribution (equivalently, using
periodic boundary conditions so that $\int dX ~\partial_X \phi(X) =0$), and
summing over $\mu=\pm 1$ leads to
\bea
H'_{2k_\nu} &=&-\frac{\cos^2[\gamma_\nu]}{2\pi} \int dx ~U(x) \cos[2k_\nu x] \nonumber\\
&&\times \int dX ~(\partial_X \phi_\nu)^2.
\eea
The integral over relative distance gives backscattering component of the potential
$U(2k_\nu)$.
Adding two contributions we find
\be
H'_{\text{intra}} = \frac{U(0) - \cos^2[\gamma_\nu] U(2k_\nu)}{2\pi} \int dx ~(\partial_x \phi_\nu)^2.
\label{ap:H'-result}
\ee
The sum of (\ref{ap:H0-intra}) and (\ref{ap:H'-result}) gives us the  result (\ref{eq:Hintra}).

\section{Perturbative Approach to Generate the Cooper term}
\label{sec:ap-C}

Expansion (\ref{eq:Z}) relies on the following operator product expansion (OPE)
\bea
 :e^{i \alpha \phi_R(\bar{z})}: :e^{i \beta \phi_R(\bar{z}')}: & = &
:e^{i [\alpha \phi_R(\bar{z}) + \beta \phi_R(\bar{z}')] }: \nonumber\\
&&\times e^{-\alpha \beta \langle \phi_R(\bar{z}) \phi_R(\bar{z}')\rangle},\nonumber\\
:e^{i \alpha \phi_L(z)}: :e^{i \beta \phi_L(z')}: &=&
:e^{i [\alpha \phi_L(z) + \beta \phi_L(z')] }: \nonumber\\
&&\times e^{-\alpha \beta \langle \phi_L(z) \phi_L(z')\rangle},
\label{ap:ope1}
\eea
where $z = u_\sigma \tau + i x$ and $\barz = u_\sigma \tau - i x$ and correlation functions of chiral bosons
are defined by
\bea
\langle \phi_L(z)\phi_L(z') \rangle &=&
-\frac{1}{4\pi}\ln (\frac{z-z'}{a_0}), \nonumber\\
\langle\phi_R(\barz)\phi_R(\barz') \rangle  &=& -\frac{1}{4\pi}\ln
(\frac{\bar{z}-\bar{z}'}{a_0}).
\label{ap:corr1}
\eea
Both (\ref{ap:ope1}) and (\ref{ap:corr1}) follow from the harmonic $S_0$,
see (\ref{eq:S0-1}).
We also employ Baker-Hausdorff formulae
\be
e^A e^B = e^B e^A e^{[A,B]}~,~e^A e^B = e^{A+B} e^{\frac{1}{2}[A,B]}
\label{ap:BH}
\ee
to convert expressions in terms of dual bosons $\varphi_\sigma, \theta_\sigma$
into those in terms of chiral bosons $\phi_R, \phi_L$,
\be
\varphi_\sigma  = \phi_L + \phi_R ~,~ \theta_\sigma  = \phi_L - \phi_R.
\label{ap:bos8}
\ee
(Note that for brevity we suppress spin index $\sigma$ on the right-hand-side of the above equation.)
Their commutation relations are given by (\ref{eq:bos2}) and (\ref{eq:bos3})
with $\nu=\nu'=\sigma$.

Series (\ref{eq:Z}) are conveniently formulated, using (\ref{ap:BH}) and $K_\sigma \to K$, in terms of
\begin{eqnarray}
&&A_{\mu \nu} (z,\barz)=
: e^{  i \sqrt{2\pi K} \mu \varphi_\sigma }e^{ i
\sqrt{\frac{2\pi}{ K}} \nu \theta_\sigma }: e^{i \mu q_0 x}=  \nonumber\\
&& = e^{i \frac{\pi}{4}(\frac{1}{K_\sigma} - K_\sigma ) -  i
\frac{\pi}{2} \mu \nu}
\times : e^{i\sqrt{2\pi}(\mu \sqrt{ K} + \frac{\nu}{\sqrt{K}})\phi_L (z)} \nonumber\\
&&\times e^ {i\sqrt{2\pi}(\mu \sqrt{ K} - \frac{\nu}{\sqrt{K}} )\phi_R (\barz)}: e^{i \mu q_0 x}.
\label{ap:amunu}
\end{eqnarray}
Indeed,
\be
\hat{H}_R = \frac{\tilde{g}_R }{4i} \int dx ~\sum_{\mu,\nu=\pm} \nu A_{\mu,\nu}.
\ee

In the second-order ($g_R^2$) term, we need to combine $A_{\mu\nu}(z,\barz)$ and $A_{\mu'\nu'}(z',\barz')$.
Using (\ref{ap:ope1}) and (\ref{ap:corr1}), we {\sl fuse} field $\phi_{R/L}$ at different points $z,z'=u_\sigma \tau' + i x'$ by setting
$z' \to z$ everywhere where this procedure does not cause divergence. This is just an OPE-based
gradient expansion. In this way we obtain
\begin{eqnarray}
&& A_{\mu\nu}(z)  A_{\mu'\nu'}(z')  =  \exp[i \frac{\pi}{2}
(\frac{1}{K}-  K)  -  i\frac{\pi}{2}(\mu \nu + \mu'\nu')] \nonumber\\
&& \times  \exp[   i \frac{\pi}{2} (\mu\mu' K  - \frac{\nu\nu'}{K} +  \mu \nu' - \nu \mu') ]  \times e^{i q_0(\mu x + \mu' x')} \nonumber\\
&& \times :\exp[i \sqrt{2\pi K}(\mu + \mu') \phi_L(z)  + i
\sqrt{\frac{2\pi}{ K}}(\nu + \nu') \phi_L(z) ] :
\nonumber\\
&& \times :\exp[i \sqrt{2\pi K}(\mu + \mu') \phi_R(\barz)  - i
\sqrt{\frac{2\pi}{ K}}(\nu + \nu') \phi_R(\barz) ] :
\nonumber\\
&&  \times \exp[-\frac{1}{2} \ln (\frac{a_0}{z-z'}) ( \mu \mu'K +
 \frac{\nu \nu'}{K} +  \mu \nu'  + \nu \mu' ) ] \nonumber\\
&& \times  \exp[-\frac{1}{2} \ln (\frac{a_0}{\bar{z}-\bar{z}'}) (
\mu \mu' K  +
 \frac{\nu \nu'}{K} -  \mu \nu'  - \nu \mu' ) ] .
\label{ap:a^2}
\end{eqnarray}
In the subsequent summation over $\mu,\mu',\nu,\nu'=\pm$ indices two
combinations, with $\mu=\mu', \nu=\nu'$ and $\mu=-\mu', \nu=-\nu'$, produce highly
irrelevant terms (of scaling dimension $\approx 4$) and are disregarded easily.
The choice $\mu=\mu'$ and $\nu=-\nu'$ produces {\sl backscattering} correction
\bea
&& A_{\mu\nu}(z)  A_{\mu, -\nu}(z') = \exp[ i \pi (K + 1/K - \mu\nu]\nonumber\\
&&\times \Big|\frac{z-z'}{a_0}\Big|^{K-1/K} e^{i \mu (\sqrt{8\pi K} \varphi_\sigma + 2 q_0 x)}.
\label{ap:B}
\eea
Finally, the choice $\mu=-\mu'$ and $\nu=\nu'$,
yields the relevant {\sl Cooper}
term,
\bea &&A_{\mu\nu}A_{-\mu\nu} =  \exp[i\pi(\mu\nu - K - 1/K)] \nonumber\\
&& \times \Big|\frac{a_0}{z-z'} \Big|^{K - 1/K}  e^{i \mu q_0 (x-x')}
e^{i \nu \sqrt{8\pi/K}~\theta_\sigma } .
\label{ap:C}
\eea
We then notice that for $K\equiv K_\sigma > 1$ (see (\ref{eq:LLparameters}) with $\gamma_F \to 0$),
the backscattering piece (\ref{ap:B})  is not singular as $z'\to z$ limit is taken and simply disappears
in this limit. Moreover, it is {\sl irrelevant} (scaling dimension $2/K_\sigma > 2$) and contains an
oscillating phase factor $2q_0 x$ which ``averages" it to zero.
The other, Cooper contribution (\ref{ap:C}) instead {\sl diverges} in this limit and thus has to be retained.

Collecting everything together, we find for the second-order correction $Z^{(2)}$ to the unperturbed $Z^{(0)} = \int e^{-S_0}$
\be
Z^{(2)} =  \frac{1}{8} \frac{\tilde{g}_R^2}{u_\sigma} f(\kappa) \int dX dT \cos[ \sqrt{8\pi/K}~\theta_\sigma(X,T)],
\ee
where $(X,T)=((x+x')/2,(\tau + \tau')/2)$  are the center of mass coordinates.
Function $f(\kappa)$, with $\kappa \equiv (K - 1/K)/2 $, is given by the integral over
the relative coordinates $(x,t) \to (x-x',\tau - \tau')$
(it comes from OPE result (\ref{ap:C}))
\begin{eqnarray}
f(\kappa) &=& \int^{\infty}_{-\infty} dx d t ~\frac{\cos[ q_0 x]}{(x^2 +
t^2)^\kappa} = -2\sqrt{\pi}\cos (\pi \kappa) \nonumber\\
&&\times q_0^{2\kappa-2}
\Gamma(2-2\kappa)\Gamma(\kappa-1/2)/\Gamma(\kappa) , \nonumber\\
&&f(\kappa\rightarrow 0)= \frac{4 \pi \kappa}{q_0^2} .
\label{ap:f}
\end{eqnarray}
Because of its convergence, the integral can be extended to infinite limits, and this was done here.
Re-exponentiating this contribution, we end up with the second-order correction $\delta S^{(2)}$ to
the free action $S_0$
\bea
\delta S^{(2)} &=& - (\eta_\uparrow\eta_\downarrow\eta_\uparrow\eta_\downarrow) \Big(\frac{\alpha_R k_F}{K_\sigma \Delta_z}\Big)^2
\frac{U(2k_F)}{(\pi a_0)^2 K_\sigma} \nonumber\\
&&\times \int dXdT  \cos[ \sqrt{\frac{8\pi}{K_\sigma}}~\theta_\sigma].
\label{ap:S2}
\eea
Note that $U(2k_F)/v_F$ comes from small-$\kappa$ limit of (\ref{ap:f})
while the ratio of spin-orbit to Zeeman energies
appears from $g_R^2/q_0^2$ combination. Finally, observe that Klein factors combine to produce
overall {\sl positive}  sign, as
$(\eta_\uparrow\eta_\downarrow)^2 = \eta_\uparrow\eta_\downarrow\eta_\uparrow\eta_\downarrow =
-\eta_\uparrow^2 \eta_\downarrow^2 = -1$,
in perfect agreement with our two-subband result in (\ref{eq:H-sigma-C}).

\section{Momentum shell RG}
\label{sec:ap-shell}

This appendix is intended to provide self-consistent description of the
standard momentum-shell RG to the problem and to highlight few
seemingly tricky technical points that arise.

Similar to the Appendix~\ref{sec:ap-C}, our starting point here are equations
(\ref{eq:Z},\ref{eq:S0-1}) and (\ref{eq:HR3}). Fields $\varphi_\sigma, \theta_\sigma$
are split into {\sl slow} (index $s$) and {\sl fast} (index $f$) components
\be
\varphi(k) = \sum_{r=s, f}\varphi_r (k)~, ~\theta(k) = \sum_{r=s, f}\theta_r (k),
\label{ap:slow-fast}
\ee
where $k=(q,\omega)$ is a two-momentum. Fast components have finite support only in the
narrow (two-) momentum shell of ``width" $d\Lambda$, and are integrated over
during the RG step. Precise shape of the momentum shell to be integrated out
will be discussed near the end, most of the calculation relies on the fact that the
area of that shell is small, $\propto d\Lambda$. Being quadratic, the free action
splits into slow $S_{0,s}$ and fast $S_{0,f}$ parts as well and,
integrating out fast modes in every order
of the perturbation expansion, one converts (\ref{eq:Z}) into a cumulant expansion for
the {\sl effective} action
\be
S_{\text{eff}} = S_{0,s} - \langle S'\rangle_f - \frac{1}{2} (\langle S'^2\rangle_f -
\langle S'\rangle_f^2) + ...,
\label{ap:cum}
\ee
where the perturbation is  $S' = \int d\tau \hat{H}_R$ and $\langle O\rangle_f$ stands for the
expectation value of $O$ evaluated with {\sl fast} action $S_{0,f}$.
The first order term just produces rescaling of the coupling constant
\be
\tilde{g}_R \to \tilde{g}'_R = \tilde{g}_R \times \exp[-\frac{1}{2}(K+1/K)\int' \frac{dk}{k}],
\label{ap:1order}
\ee
where $\int'$ denotes integration over the fast component support area. The factor
in front of the integral is just the scaling dimension of the spin-orbit operator $\hat{H}_R$,
$\frac{1}{2}(K+1/K) \approx 1$. Rescaling space-time {\sl back} produces additional
factor of $2$ in the exponent in (\ref{ap:1order}): $\frac{1}{2}(K+1/K) \to 2-\frac{1}{2}(K+1/K)$.
This, using $\int' \frac{dk}{k} = d\Lambda/\Lambda = d\ell$, leads finally to the standard
first-order RG equation for the running coupling constant,
$d \tilde{g}_R/d\ell = (2 -\frac{1}{2}(K+1/K)) \tilde{g}_R$. Note that Klein
factors $\eta_{\uparrow,\downarrow}$ do not affect the scaling in any way.
The second-order contribution contains Cooper and backscattering terms,
\bea
&&\delta S^{(2)} = \frac{(\tilde{g}'_R)^2}{8} \int_{1}\int_{2}
[B(r_{12}) -1] \cos[\sqrt{2\pi K} (\varphi_{s,1}-\varphi_{s,2}) \nonumber\\ &&+ q_0 (x_1 - x_2)]
\cos[\sqrt{\frac{2\pi}{K}} (\theta_{s,1}+\theta_{s,2})] - [B^{-1}(r_{12}) -1] \nonumber\\
&&\times \cos[\sqrt{2\pi K} (\varphi_{s,1} +\varphi_{s,2}) + q_0 (x_1 + x_2)] \nonumber\\
&& \times\cos[\sqrt{\frac{2\pi}{K}} (\theta_{s,1}-\theta_{s,2})],
\label{ap:2order}
\eea
where $1(2)$ are short-hand notations for $\vec{r}_{1,2} = (x_{1,2},\tau_{1,2})$,
$\vec{r}_{12} = \vec{r}_1 - \vec{r}_2$,
and, for example, $\varphi_{s,1} \equiv \varphi_s(x_1,\tau_1)$. Here
\be
B(r_{12}) = \exp\{(K-1/K) \int' \frac{d^2\vec{k}}{2\pi} \frac{ \cos[\vec{k}\cdot\vec{r}_{12}]}{k^2}\}
\label{ap:B12}
\ee
appears from integrating out {\sl fast} components of bosonic fields.
Performing gradient expansion of (\ref{ap:2order}) we find two contributions
\bea
&&\delta S^{(2)}_{\text{C}} = \frac{(\tilde{g}'_R)^2}{8} B_C \int d^2\vec{R}
\cos[\sqrt{\frac{8\pi}{K}} \theta_s(R)] \label{ap:2order-C}\\
&&\delta S^{(2)}_{\text{bs}} = \frac{-(\tilde{g}'_R)^2}{8} B_{\text{bs}} \int d^2\vec{R} \cos[\sqrt{8\pi K} \varphi_s(R) + \nonumber\\ &&+ 2 q_0  X], \label{ap:2order-bs}
\eea
where the center-of-mass coordinates are as usual
$\vec{R}=(X,T)=(\vec{r}_1 + \vec{r}_2)/2$.
Their amplitudes are given by
\bea
&& B_C = \int d^2\vec{r} ~[B(r) -1] \cos[q_0 x], \label{ap:2order-BC}\\
&& B_{\text{bs}} =  \int d^2\vec{r} ~[B^{-1}(r) -1], \label{ap:2order-Bbs}
\eea
and differ by the absence of {\sl cosine} factor in the second, backscattering-related amplitude
(\ref{ap:2order-Bbs}).
The final step is to {\sl expand} $B(r)$ as according to its definition (\ref{ap:B12})
the expression in the exponential is proportional to small $d\Lambda$ \cite{thanksLB}.
One then finds that (\ref{ap:2order-BC}) is proportional to the product of the integral over
the relative coordinate, $\int d^2\vec{r}$ {\sl and} the integral over {\sl fast} mode support, $\int' d^2 \vec{k}$,
coming from (\ref{ap:B12}). The coordinate integral is performed first to obtain
\be
B_C = \pi (K-1/K) \int' d\omega dq \frac{\delta(\omega)[\delta(q-q_0) + \delta(q+q_0)]}{\omega^2 + q^2}.
\label{ap:BC}
\ee
We finally argue that magnetic field, which determines $q_0$ (\ref{eq:HR3}), breaks the
symmetry between space $x$ and time $\tau$ and this allows us to {\sl choose}
an asymmetric prescription for $\int'$. Namely, we integrate over {\sl all} frequencies
while the $q$-integration is restricted to the $\pm |\Lambda - \Lambda'|$ interval.
This gives
\bea
&&B_C =\left\{
\begin{array}{cc}
2\pi (K-1/K)/q_0^2 & \text{if} ~q_0 \in (\Lambda - d\Lambda, \Lambda)~~~\\
0 & \text{otherwise}.
\end{array}
\right.
\label{ap:BCfinal}
\eea
Note that the result does not contain $d\ell = d\Lambda/\Lambda$ which
serves to emphasize its meaning as a fluctuation-degenerated {\sl initial value}
of the Cooper term. Let us now see what this approach predicts for the
backscattering term (\ref{ap:2order-Bbs}).
Expanding $B(r)$ as in (\ref{ap:BC}) we get
\bea
&& B_{\text{bs}} = - 2\pi (K-1/K) \int' d\omega dq \frac{\delta(\omega)\delta(q)}{\omega^2 + q^2}\nonumber\\
&&+ \frac{1}{2} (K-1/K)^2 \int' d\omega dq \frac{1}{(\omega^2 + q^2)^2} + ...
\label{ap:Bbs}
\eea
While the first term is clearly zero, the second, which originates from the second order expansion
of (\ref{ap:B12}), is clearly finite for any shape of the {\sl fast} modes support. This,
{\sl quadratic} in $U(2k_F)/v_F \ll 1$ result, is in agreement with a non-singular structure
of the similar correction found during the real-space calculation in (\ref{ap:B}).
Note finally that in the absence
of magnetic field ($q_0=0$) this scheme predicts similar in structure (but different in signs) corrections
to the Cooper (\ref{ap:2order-C}) and backscattering (\ref{ap:2order-bs}) terms, in agreement
with the result of real-space OPE calculations, see Chap.20 in Ref.[~\onlinecite{gnt-book}]
and papers \onlinecite{nersesyan,yakovenko}.

Now we return to the original problem with finite $q_0 \neq 0$.
Combining (\ref{ap:BCfinal}) with (\ref{ap:2order-C}) we have
\be
\delta S^{(2)}_{\text{C}} = -\Big(\frac{\alpha_R k_F}{\Delta_z K_\sigma}\Big)^2 \frac{U(2k_F)}{4 (\pi a_0)^2}
\int d^2\vec{R} \cos[\sqrt{\frac{8\pi}{K}} \theta_s(R)],~~~~~~~
\label{ap:C-final}
\ee
which agrees with (\ref{ap:S2}) and (\ref{eq:H-sigma-C}) in everything {\sl but sign}!
That sign comes from the Klein factors in $\tilde{g}_R$ (note that at this stage the difference
between $\tilde{g}_R$ and $\tilde{g}'_R$ is of higher order in $d\ell$ and not important)
and is a consequence of the identity $(\eta_\uparrow \eta_\downarrow)^2 = -1$.
This puzzling discrepancy between (\ref{eq:H-sigma-C},\ref{ap:S2}) and (\ref{ap:C-final})
is worth figuring out in detail.

We note that within the functional integral approach, which is the framework for the
momentum shell RG described here, {\sl all} information on commutation relations of
dual fields $\varphi$ and $\theta$ is contained in $(-i\partial_\tau \varphi \partial_x \theta)$
term in the bare action (\ref{eq:S0-1}). This is nothing but field-theoretic version of
the canonical $p \dot{x}$ term in quantum mechanics \cite{wen-book}.
It identifies $\varphi$ as a ``coordinate" and $\partial_x \theta$ as a ``momentum":
$[\varphi(x), \partial_{x'} \theta(x')] = i \delta(x-x')$. While fully consistent with our basic
commutation relation (\ref{eq:bos7}), this canonical bracket {\sl does not} contain
information on the non-trivial commutation relation (\ref{eq:bos2}) of chiral right and
left bosons. Indeed, it is a simple exercise to see that replacing (\ref{eq:bos2}) with
{\sl commuting} chiral bosons $[\tilde{\phi}_R, \tilde{\phi}_L]=0$ changes our (\ref{eq:bos7})
into $[\tilde{\varphi}(x), \tilde{\theta}(x')]= i ~\text{sign}(x'-x)/2$ which still satisfies
``coordinate-momentum" identification for the pair $\tilde{\varphi}$ and $\partial_x \tilde{\theta}$.
To put things differently, an analog of bosonic action (\ref{eq:S0-1}) expressed
in terms of ``tilded" fields $\tilde{\varphi}$ and
$\tilde{\theta}$ is identical to the current one in terms of our $\varphi$ and $\theta$ fields
satisfying equations (\ref{eq:bos2}, \ref{eq:bos3}) and (\ref{eq:bos7}).
This simply means: bosonic action $S_0$, (\ref{eq:S0-1}), does {\sl not}
enforce anticommutation relations between {\sl right} $R$ and {\sl left} $L$ moving
fermions (\ref{eq:bos1}). This shortcoming of bosonic functional integral is well known,
see for example Appendix C in Ref.~\onlinecite{giamarchi-book},
 and several ``fixes" were proposed in the literature.
Our approach consists in enlarging the role of Klein factors: instead of
(\ref{eq:bos1}) we bosonize fermions here as
\bea
R_{s} = \frac{\kappa_{Rs}}{\sqrt{2\pi a_0}}  e^{i\sqrt{4\pi}
\tilde{\phi}_{Rs} },\ \
 L_{s} = \frac{\kappa_{Ls}}{\sqrt{2\pi a_0}}
e^{-i\sqrt{4\pi} \tilde{\phi}_{Ls} }.
\label{ap:bos8}
\end{eqnarray}
The Klein factors $\kappa_{R/L,s}$ now carry {\sl double} index: chirality ($R$ or $L$)
and spin ($s$). Even though the chiral bosons $\tilde{\phi}_{Rs}$ and
$\tilde{\phi}_{Ls}$ now {\sl commute}, the anticommutation of $R_s$ and $L_s$
is enforced by the Klein factors:
$\{\kappa_{\lambda,s},\kappa_{\lambda',s'}\}=2 \delta_{\lambda,\lambda'} \delta_{s,s'}$,
with $\lambda=R/L$.

As a result of the proposed modification the spin-orbit term (\ref{eq:HR3}) has to be
modified. It is convenient, following Ref.~\onlinecite{marston}, to introduce
\be
\hat{\Gamma} = \kappa_{R\uparrow}  \kappa_{R\downarrow}  \kappa_{L\uparrow}  \kappa_{L\downarrow} .
\label{ap:gamma}
\ee
This product satisfies $\hat{\Gamma}^2 = 1$, from where it follows that its eigenvalues are
$\Gamma = \pm 1$. One also checks that $[\hat{\Gamma}, \kappa_{R\uparrow}  \kappa_{R\downarrow}] = 0$
and $\kappa_{R\uparrow}  \kappa_{R\downarrow} \hat{\Gamma} = - \kappa_{L\uparrow}  \kappa_{L\downarrow}$.
These properties allow us to represent $\hat{H}_R$ as
\bea
&&\hat{H}_R = G_R  \int dx \{\frac{(1+\hat{\Gamma})}{2} \cos[\sqrt{2\pi K} \tilde{\varphi} + q_0 x]
\sin[\sqrt{\frac{2\pi}{K}}\tilde{\theta}] \nonumber\\
&& - \frac{(1-\hat{\Gamma})}{2} \sin[\sqrt{2\pi K} \tilde{\varphi} + q_0 x] \cos[\sqrt{\frac{2\pi}{K}}\tilde{\theta}] \},
\label{ap:HR4}\\
&&G_R = \frac{2 \alpha_R k_F }{\pi a_0} (i \kappa_{R\uparrow}  \kappa_{R\downarrow}).
\label{ap:GR}
\eea
Finally one observes that $\hat{\Gamma}$ {\sl commutes} with the Hamiltonian of the problem,
which implies that its eigenvalues represent integrals of motion. That is, the choice $\Gamma = +1$
or $\Gamma = -1$ is the {\sl gauge} choice, and one can replace operator
$\hat{\Gamma}$ in (\ref{ap:HR4}) by its eigenvalue $\Gamma$.
Comparing (\ref{ap:HR4}) with our original (\ref{eq:HR3})
we observe that the latter corresponds to $\Gamma=+1$ gauge: with $\Gamma=+1$, equation
(\ref{ap:HR4})  transforms into (\ref{eq:HR3}) by replacing
$(\tilde{\varphi},\tilde{\theta}) \to (\varphi_\sigma, \theta_\sigma)$.
Repeating steps that led us to (\ref{ap:C-final}) we arrive at
\be
\delta \tilde{S}^{(2)}_{\text{C}} = \frac{\pi}{4} \Big(\frac{G_R}{q_0}\Big)^2 (K-1/K)
\int d^2\vec{R} \cos[\sqrt{\frac{8\pi}{K}} \tilde{\theta}].
\label{ap:C-tilde}
\ee
The only, but key, difference with (\ref{ap:C-final}) is that now
$(i \kappa_{R\uparrow}  \kappa_{R\downarrow})^2 = - (\kappa_{R\uparrow}  \kappa_{R\downarrow})^2 =
\kappa_{R\uparrow}^2 \kappa_{R\downarrow}^2 = +1$, which implies that
the amplitude of the Cooper term is {\sl positive} in (\ref{ap:C-tilde}), in final
agreement with our previous and independently derived results in (\ref{eq:H-sigma-C},\ref{ap:S2}).
This amusing exercise illustrates the importance of Klein factors, and, more generally,
of preserving correct (anti)commutation relations when implementing convenient
but tricky bosonization formalism. We conclude by noting that results of the other gauge choice,
$\Gamma=-1$ in (\ref{ap:HR4}), while equivalent in principle to the one made above,
are most conveniently understood as following from the {\sl global} shift of bosonic fields:
$\tilde{\varphi} \to \varphi + \sqrt{\pi/(8K)}$ and $\tilde{\theta} \to \theta + \sqrt{\pi K/8}$.
This shift must be made in {\sl all} bosonized expressions: it changes overall sign in (\ref{ap:C-tilde})
but this is ``compensated" by the effect of the global shift described here.
As a result, the physical meaning of the Cooper instability as that of the SDW$_x$ instability
remains intact.

\section{Perturbative Approach to Generate the Impurity Term }
\label{sec:ap-imp}

The purpose of this section is to show that inter-subband contribution to impurity potential,
second term in Eq.~\ref{eq:Vback}, can be obtained as a result of
interference between local impurity backscattering (first term in (\ref{eq:Vback})) and bulk
spin-orbit (\ref{eq:HR2}). Being interested in the interference between two single-particle
terms, we perform the calculation directly in fermion fields and specialize to the limit
$\Delta_z \gg E_{\text{s-o}}$ ($\gamma_F \to 0$) for simplicity.

The lowest order in the perturbative expansion involving  these two terms is obtained by the
following correction to partition function (compare with (\ref{eq:Z}))
\bea
\delta Z_{\text{imp}} &=& \frac{1}{2} \int e^{-S_0}  \int d\tau d\tau' \hat{H}_R(\tau) V^B(\tau') \nonumber\\
&&\to \int e^{-S_0} \int d\tau ~\delta V^B,
\label{ap:Z-imp}
 \eea
 where the second line identifies correction to the impurity backscattering term we are after.
 Here $S_0$ is the action of two Zeeman-split $\{\uparrow,\downarrow\}$ subbands, and
 $V^B$ is the intra-subband
term due to the impurity, located at $x=0$,  given by
\be
V^B = V_0 \sum_{s=\uparrow, \downarrow}\Big( R^\dagger_s L_s + L^\dagger_s R_s \Big)_{x=0}.
\label{ap:VB}
\ee
The spin-orbit term $ \hat{H}_R$ in the presence of magnetic field (so that $k_\uparrow - k_\downarrow = \delta k_F$)
reads
\be
\hat{H}_R = \alpha_R k_F \int dx \sum_{s=\uparrow, \downarrow} \Big(R^\dagger_s R_{-s} e^{-i s \delta k_F x} -
L^\dagger_s L_{-s} e^{i s \delta k_F x}\Big),
\label{ap:HR5}
\ee
where we neglected terms $\propto \delta k_F/k_F = \Delta_z/E_F \ll 1$.

We calculate (\ref{ap:Z-imp}) by fusing fermions with like spin index, that is by making the replacement
$\Psi_{R/L s}(x,\tau) \Psi^\dagger(x',\tau') \to G_{R/L}(x-x',\tau-\tau')$ where possible. Here $G_{R/L}$ stands
for fermions Green's functions. To lowest (zeros) order in the interaction these
are given by the {\sl free} fermion Green's functions \cite{sfb}
\bea
G_R(x,\tau) &=& \langle\Psi_{Rs}(x,\tau)\Psi^\dagger_{Rs'}(0)\rangle = \frac{\delta_{s,s'}}{2\pi (v\tau - ix)}\nonumber\\
G_L(x,\tau) &=& \langle\Psi_{Ls}(x,\tau)\Psi^\dagger_{Ls'}(0)\rangle = \frac{\delta_{s,s'}}{2\pi (v\tau + ix)}.
\label{ap:GRL}
\eea
In this way we find
\bea
\int d\tau ~\delta V^B &=& \frac{\alpha_R k_F V_0}{4\pi} \int d\tau \Big( (R^\dagger_\uparrow L_\downarrow -
L^\dagger_\uparrow  R_\downarrow) P_1 \nonumber\\
&&+ (R^\dagger_\downarrow L_\uparrow - L^\dagger_\downarrow R_\uparrow) P_2 \Big),
\eea
where
\bea
P_1 &=& \int dx dt (\frac{e^{-i\delta k_F x}}{\barz} + \frac{e^{i\delta k_F x}}{z}) \\
P_2 &=&\int dx dt (\frac{e^{i\delta k_F x}}{\barz} + \frac{e^{-i\delta k_F x}}{z}),
\eea
and $z=v_F t + ix, \barz = v_F t - ix$ as usual. These integrals are easily calculated
\be
P_1 = - P_2 = \frac{4\pi}{\delta k_F v_F^2}.
\ee
Expressing $\delta k_F = \Delta_z/v_F$ we obtain the correction
\be
\delta V^B = \frac{\alpha_R k_F V_0}{\Delta_z}  \Big(R^\dagger_\uparrow L_\downarrow -
R^\dagger_\downarrow L_\uparrow + L^\dagger_\downarrow R_\uparrow - L^\dagger_\uparrow  R_\downarrow\Big).
\label{ap:delta-VB}
\ee
We observe that $\delta V^B$ is {\sl odd} under spatial inversion ${\cal P}$
(with respect to impurity location)
when $x\to -x$ and right- and left-movers get interchanged, $R_s \leftrightarrow L_s$.
Of course, it {\sl must} be odd under ${\cal P}$ as it is obtained from
fusing {\sl even} \eqref{ap:VB} and {\sl odd} \eqref{ap:HR5} in ${\cal P}$ terms.
The oddness of \eqref{ap:delta-VB} is the reason for the relative minus signs in
this equation.

Bosonization of \eqref{ap:delta-VB}, following Sec.~\ref{sec:bosonization}, results in
\be
\delta V^B = \frac{2\alpha_R k_F V_0}{\Delta_z}
\frac{i \eta_\uparrow \eta_\downarrow}{\pi a_0} \sin[\sqrt{2\pi}\varphi_\rho]
\cos[\sqrt{2\pi} \theta_\sigma] ,
\label{ap:delta-VB-bos}
\ee
which confirms our previous result (\ref{eq:rho-inter},\ref{eq:Vback}).
The generated impurity potential describes {\sl intra}-subband impurity backscattering.

The absence of the potentially
dangerous term with $\sin[\sqrt{2\pi} \theta_\sigma]$ in place of
$\cos[\sqrt{2\pi} \theta_\sigma]$ in \eqref{ap:delta-VB-bos}  is now clear:
such a term would require no minus signs in \eqref{ap:delta-VB}
which is forbidden by the oddness of $\delta V^B$ under inversion ${\cal P}$.

\end{document}